\documentclass[amsmath,amssymb,aps,letterpaper,prx,twocolumn,longbibliography,superscriptaddress]{revtex4-1}

\usepackage{graphicx}
\usepackage{color}
\usepackage{enumitem}

\usepackage{picture}
\usepackage{calc}

\usepackage{subfigure}
\usepackage{dsfont}
\usepackage{color}

\usepackage[utf8x]{inputenc}

\usepackage{tikz}
\usepackage[english]{babel}
\usepackage[T1]{fontenc}
\usepackage{indentfirst}
\usepackage{amsmath}
\usepackage{amssymb}
\usepackage{amsthm}
\usepackage{proof}
\usepackage{eufrak}
\usepackage{graphicx}
\usepackage{psfrag}

\usepackage[normalem]{ulem}

\allowhyphens

\newcommand{\egesz}{\mathbb{Z}}

\newcommand{\ordo}{\mathcal{O}}

\newcommand{\pn}{{\boldsymbol p}_N}

\newcommand{\pp}{{\boldsymbol p}}
\newcommand{\kk}{{\boldsymbol k}}
\newcommand{\km}{{\boldsymbol k}_M}

\newcommand{\emp}{\mathbb{E}}

\newcommand{\abs}[1]{|#1|}

\newcommand{\vev}[1]{\left\langle #1 \right\rangle}
\newcommand{\ket}[1]{{\left|#1\right\rangle}}
\newcommand{\bra}[1]{{\left\langle #1\right|}}

\newcommand{\skalarszorzat}[2]{{\langle #1 | #2 \rangle}}

\newcommand{\uc}{p}
\newcommand{\ub}{k}
\newcommand{\buc}{\boldsymbol p}
\newcommand{\bub}{\boldsymbol k}


\usepackage{ifpdf}

\ifpdf
\usepackage{epstopdf}
\usepackage[pdftex,colorlinks,urlcolor=blue,citecolor=blue,linkcolor=blue]{hyperref}
\else
\usepackage[hypertex,colorlinks,urlcolor=blue,citecolor=blue,linkcolor=blue]{hyperref}
\fi
\begin{document}
%\numberwithin{equation}{section}

\title{
  An integrable spin chain with Hilbert space
  fragmentation
 % , fracton-like excitations,
  and solvable real time dynamics
  % many body scars and TTbar deformation
}

\author{Bal\'azs Pozsgay}
\affiliation{MTA-ELTE “Momentum” Integrable Quantum Dynamics Research Group, Department of Theoretical Physics, Eötvös Loránd University}
\author{Tam\'as Gombor}
\affiliation{MTA-ELTE “Momentum” Integrable Quantum Dynamics Research Group, Department of Theoretical Physics, Eötvös Loránd University}
\affiliation{Holographic QFT Group, Wigner Research Centre for Physics, Budapest, Hungary}
\author{Arthur Hutsalyuk}
\affiliation{MTA-ELTE “Momentum” Integrable Quantum Dynamics Research Group, Department of Theoretical Physics, Eötvös Loránd University}
\author{Yunfeng Jiang}
\affiliation{Department of Theoretical Physics, CERN, 1 Esplanade des Particules, Geneva 23, CH-1211, Switzerland.}
\affiliation{Shing-Tung Yau Center and School of Physics, Southeast University, Nanjing 210096, China.}
\author{Levente Pristy\'ak}
\affiliation{Department of Theoretical Physics, Budapest University of Technology and Economics}
\affiliation{MTA-ELTE “Momentum” Integrable Quantum Dynamics Research Group, Department of Theoretical Physics, Eötvös Loránd University}
\author{Eric Vernier}
\affiliation{CNRS \& LPSM, Universit\'e de Paris, place Aur\'elie Nemours, 75013 Paris, France}

\begin{abstract}
  We revisit the so-called folded XXZ model, which was treated earlier by two independent research groups. We argue
  that this  spin-1/2 chain is one of the simplest quantum integrable models, yet it has quite remarkable physical
  properties. The particles have constant scattering
  lengths, which leads to a simple treatment of the exact spectrum and the dynamics of the system. The Hilbert space of
  the model is fragmented, leading to exponentially large degeneracies in the spectrum,
such that the exponent depends on the particle content of
  a given state.
  We provide an alternative derivation of the Hamiltonian and the conserved charges of the model, including a new
 interpretation  of the so-called ``dual model'' considered earlier. We also construct a non-local map that
 connects the model with the Maassarani-Mathieu spin chain, also known as the $SU(3)$ XX model. We consider the exact
 solution of the model with periodic and open boundary conditions, and also derive multiple descriptions of the exact thermodynamics of the model. We consider quantum quenches of different types. In one class of problems the dynamics can be treated
 relatively easily: we compute an example for the real time dependence of a local observable. In another class of
 quenches the degeneracies of the model lead to the breakdown of equilibration, and we argue that they can lead to
 persistent oscillations. We also discuss connections with the $T\bar T$ and hard rod deformations known from Quantum Field Theories.
 \end{abstract}

\maketitle

\section{Introduction}

Quantum integrable models are special many body systems that allow for an exact solution, at least for certain observables and
in certain situations. They possess a large number of extra conservation laws, which constrain their dynamics, eventually
allowing their solvability \cite{caux-integrability}.
Consequently, they have been used fruitfully in the past to compute ground-state or thermal properties in various
situations, with applications ranging from condensed matter to high-energy physics \cite{essler-condmat-applications,fermi-gases-experimental-review,ads-cft-review}.

More recently, new challenges have been raised by the study of the non-equilibrium dynamics for such systems.
A first set of questions is concerned with the relaxation dynamics of physical observables, for instance after a quantum quench \cite{cardy-calabrese-quench1}. While it is now well-established that integrable models do not thermalize and instead equilibrate at late times to steady states described by the Generalized Gibbs Ensemble (VEG) \cite{essler-fagotti-quench-review}, the exact computation of the finite time dynamics using the traditional methods of integrability  has remained a very difficult task, and to this date only few results are available, associated with some very specific observables \cite{sajat-nonanalytic} or particular models (see below).
Another question is that of transport properties, which
can be described by the recent theory of Generalized Hydrodynamics (GHD)
\cite{doyon-ghd,jacopo-ghd}. Even though GHD is largely successful, it relies on some assumptions which have not yet
been proven in general \cite{doyon-gge,doyon-hydro-proj}. Some elements of the theory can be considered proven, for
example key statements about the mean values of current operators (see the review \cite{sajat-currents-review}), but
it would be desirable to rigorously prove more aspects of the theory. It was shown in the remarkable work
\cite{granet-essler-ghd} that
certain statements of GGE and GHD can be checked in the Lieb-Liniger model in a large coupling expansion; up to date
this result is one of the most convincing analytic checks of GGE and GHD (for closely related works see
\cite{spyros-ghd-1,spyros-ghd-2,milos-axel-javulo,doyon-gge,doyon-hydro-proj}).  Nevertheless there remains a need for simple toy models, which have
genuine interactions in them, and which can lead to exact proofs of the GHD predictions.

Therefore, the question that motivates the present work is the following: What are the simplest integrable models? By
this we mean models that have genuine interactions (in contrast with those related with free theories as for example the
famous XX chain), but which are nevertheless simple enough to bypass some of the difficulties presented by generic
integrable models.
 We should note that the more
conventional problem of equilibrium physics can also benefit from simple toy models, because the computation of
correlation functions is a notoriously difficult problem in Bethe Ansatz solvable models (see for example
\cite{kluemper-goehmann-finiteT-review,takacs-szecsenyi-2p,karol-hab,sajat-ujLM,granet-essler-2}).

To address this question, let us discuss some simple models that already appeared in the literature.

One of the most important
examples is the hard rod gas in the continuum; for the classical hard rod gas see
\cite{nagamiya-hardrod,rubin-hardrod,sutherland-hardrod,hard-rod-gas,hard-rod-gas2,doyon-spohn-hardrod}, for the quantum case see \cite{doyon-cardy-ttbar,yunfeng-hardrod-1,yunfeng-hardrod-2}. In these
models the fundamental particles have finite and fixed widths, and the speed of wave propagation
depends on this width through the modification of the free space between the particles. Alternatively, the scattering in
these models is such that it leads to a fixed displacement given by the hard rod length. In the classical model the
emergence of hydrodynamics could be proven for a wide range of initial conditions in \cite{hard-rod-gas}.

Another important example is the Rule 54 model, which is sometimes claimed to be the simplest interacting integrable model. It is
a classical deterministic cellular automaton originally developed in \cite{rule54}, which has been an object of interest
in the last couple of years (see the review \cite{rule54-review}). There are two fundamental particles in the model: the
left- and right-movers, and their scattering leads to a constant displacement of 1 lattice unit. Despite its simplicity,
the model shows many distinguishing features of generic interacting many-body systems, such as relaxation at late times
and coexistence of ballistic and diffusive transport \cite{rule54-transport}. The
exact real time evolution in this system was computed for special initial states in the recent work
\cite{katja-bruno-lorenzo-rule54}, which quite remarkably does not use the standard methods of integrability. Instead,
it is built on methods developed for ``dual unitary quantum gate'' models, see \cite{dual-unitary-0,prosen-dual1,dual-unitary-2}. The
dual unitary models are not integrable in the traditional sense, nevertheless they allow for exact solutions
\cite{prosen-dual1,dual-unitary-2}.

A less well known model which nevertheless belongs in this list is the so-called phase model, which arises as the
$q\to\infty$ limit of the $q$-bison lattice model.
The model was first studied in   \cite{qbozon-bog-bullough1,qbozon-bog-bullough2,qbozon-bog-bullough3}, and later also
in  \cite{qboson-izergin-kitanine-bog,qboson-keiichi,qboson-bog1,qboson-symmetric,qboson-bog2,qboson-bog3} with the
focus being on
equilibrium correlation functions and relations with dominatrix and the theory of symmetric functions.
Real time dynamics in the model was first investigated in \cite{sajat-qboson}, where it was rigorously shown for a certain
quench that the system equilibrates to the GGE prediction. This was achieved by computing the exact time dependence of a
local observable and comparing its asymptotic value to the GGE average; to our best knowledge this was the first example
for such an exact computation in a genuinely interacting case.
Afterwards further quench problems were considered in
  \cite{sajat-q2}, together with a special non-local connection with the XX chain. Although it is not explicitly stated
  in \cite{sajat-qboson,sajat-q2}, the scattering of fundamental bosons in this model is such that the trajectories of
  the particles get displaced by one lattice site.

A common property of these simple models is that they can be considered as deformations of free models, although the
deformation is highly non-local. In the case of the hard-rod gas the deformation in question is in the class of the
famous $T\bar T$-deformations, as discussed in \cite{doyon-cardy-ttbar,yunfeng-hardrod-1}. For the $q$-boson the
deformation is given by the non-local mapping to the XX chain. In the case of the Rule 54 model such an explicit
deformation has not yet been worked out, but connections with the $T\bar T$-deformations were already pointed out in
\cite{rule54-ttbar}.

More recently another relatively simple model was studied in \cite{fracton1} and later in \cite{folded1,folded2}, where
it was called the ``folded XXZ model''; the first appearance of the Hamiltonian was apparently in
\cite{maurizio-coreprere}, where it was obtained as an effective Hamiltonian. In
 \cite{folded1,folded2,maurizio-coreprere} the model was derived by considering the large $\Delta$ behaviour of the famous XXZ chain (for closely
related works see \cite{pronko-abarenkova-ising-limit,xxz-triple-point}). The model has a four site Hamiltonian and a very special dynamics: it has a sector which is
equivalent to the so-called constrained XXZ model at the free fermion point
\cite{constrained1,constrained2,constrained3}, where the particles have genuine interactions originating from a hard rod
constraint. However, the full Hilbert space of the model is much larger due to the presence of an additional type of
particle (which can be called domain wall, or DW). The DW's are not dynamical: in the absence of particles they lead
to frozen configurations and exponentially degenerate energy levels; this phenomenon was interpreted as ``Hilbert space
fragmentation'' in  \cite{fracton1} and it was also discussed in \cite{folded1}. It is also important that the DW's interact with the particles and thus
they affect the dynamics of the model. A particle can be considered as a bound state of two DW's, thus becoming
dynamical; this phenomenon bears some similarities with the so-called fractonic excitations
\cite{fracton-review1,fracton-review2}.

The exact coordinate Bethe Ansatz solution of the model was presented in \cite{folded1,folded2} and connections were found
to other existing models in the literature. However, the algebraic origin of the model and its conserved charges
was not clarified completely, and a number of interesting features of the model were not yet explored.

In this paper we contribute by a completely independent derivation of the ``folded XXZ model'' and its Bethe Ansatz solution
\footnote{We discovered the model independently, but later we noticed that it was already treated (also independently) in
 \cite{fracton1} and in \cite{folded1,folded2}.}.  Also, we point out new connections to existing models in the literature. We
explain that the model is in a special class of systems, which have constant scattering lengths.
This class includes the hard rod gas, the Rule 54 model, and the
phase
model mentioned above. We also discuss relations with the $T\bar{T}$ deformation, and in an independent line of computation we present
an exact solution of a quantum quench, mirroring some of the computations of \cite{sajat-qboson,sajat-q2}.
The results show that this special integrable model sits at the intersection of three very active yet seemingly distant fields:
out-of-equilibrium dynamics, $T\bar{T}$ deformation and fracton systems. This makes the model truly unique and highly
interesting. The exact solution of the model might shed lights on the possible exciting connections between these
research areas.

\section{The Model}

We consider a spin-1/2 chain with periodic boundary conditions.
The Hamiltonian is \cite{fracton1,folded1,folded2}
\begin{equation}
  \label{H}
  H=Q_4+hQ_1+\mu Q_2.
\end{equation}
Here $Q_1$, $Q_2$ and $Q_4$ are mutually commuting extensive operators
that are given below, and  $h$ and $\mu$ are to be understood as a magnetic field and a chemical potential.

The kinematical part of the Hamiltonian is
\begin{equation}
  \label{Q4}
  Q_4=-\frac{1}{4}\sum_{j=1}^L (1+\sigma^z_j\sigma^z_{j+3})( \sigma^+_{j+1}\sigma^-_{j+2}+ \sigma^-_{j+1}\sigma^+_{j+2}).
\end{equation}
This is a 4-site operator, which generates a spin exchange between neighbouring sites, controlled by the state of two
further neighbours. To be precise, the exchange between sites $j+1$ and $j+2$ has amplitude -1/2 if the state of the sites
$j$ and $j+3$ is the same, and zero amplitude if it is different. The numerical pre-factor is added for later
convenience.

The remaining charges are
\begin{equation}
  \label{charges12}
  \begin{split}
       Q_1&=\sum_{j=1}^L \frac{1}{2}(1-\sigma^z_j),\\
    Q_2&=\sum_{j=1}^L\frac{1}{2} (1-\sigma^z_j\sigma^z_{j+1}).\\
  \end{split}
\end{equation}
$Q_1$ can be interpreted as particle number and $Q_2$ as a domain wall number. It can be checked that the three
operators are mutually commutative.

The Hamiltonian \eqref{H} first appeared in \cite{maurizio-coreprere} as an effective Hamiltonian. Afterwards it was
studied independently in \cite{fracton1}, although that work only treated it with open boundary
conditions. This case will be studied separately in Section \ref{sec:boundary}. Afterwards the model also appeared in
\cite{folded1,folded2}, where an exact solution was given. To be precise, the papers \cite{folded1,folded2} treated an
equivalent dual model, which is discussed later in our Section \ref{sec:bond}. These papers focused on the periodic case
and they did not work out the solution for $h\ne 0$.
There is considerable overlap between the results of \cite{folded1,folded2} and our work. We choose to present a
complete treatment of the model from our point of view, meanwhile also explaining what is new and what was already given
in  \cite{fracton1} and/or \cite{folded1,folded2}.

\subsection{Relation to the constrained XXZ model}

\label{sec:constrained}

There are various ways of writing the Hamiltonian, and there are multiple connections to known integrable models in the literature.
One of these connections is with the constrained XXZ model, which describes a spin chain with XXZ type interaction, where
particles have a finite ``width'' $l\in\egesz^+$.
The model was treated in a number of works
\cite{constrained1,constrained2,constrained3,pronko-abarenkova-ising-limit,xxz-triple-point}, and its Hamiltonian is as
follows. Let us choose a convention that the down spins are interpreted as particles. Then the model is given by
\begin{equation}
  \label{cXXZ}
  H_{c}=\sum_{j=1}^L \mathcal{P}_l \left[
\sigma^x_{j}\sigma^x_{j+1}+\sigma^y_{j}\sigma^y_{j+1}+\Delta \sigma^z_{j}\sigma^z_{j+l+1}\right]
  \mathcal{P}_l,
\end{equation}
where $\mathcal{P}_l$ is a projector onto the states of the Hilbert
space where there are at least $l$ up spins between
two down spins. In other words, the linear space selected by $\mathcal{P}_l$ describes particles that have a ``width''
$l+1$.

It is known that the constrained XXZ model is integrable for every $l$ and $\Delta$, its Bethe Ansatz solution can be
found in \cite{constrained1,constrained2}. Furthermore, an algebraic treatment of its integrability properties was
given in \cite{constrained3}, where it was related to a vertex model with long range interactions. This implies the
existence of an infinite family of commuting conserved charges for the model.

To establish a relation with our model we write
\begin{equation}
  1+\sigma^z_j\sigma^z_{j+3}=2(E^{-}_{j}E^-_{j+3}+E^{+}_{j}E^{+}_{j+3}),
\end{equation}
where $E^{\pm}_j$ are projection operators to the up/down spins on site $j$. Then
we can see that $Q_4=Q_4^++Q_4^-$, where
\begin{equation}
  Q_4^\alpha=-\frac{1}{2} \sum_{j=1}^L E^\alpha_j E^\alpha_{j+3}
( \sigma^+_{j+1}\sigma^-_{j+2}+ \sigma^-_{j+1}\sigma^+_{j+2}).
\end{equation}
The two operators $Q_4^\alpha$ are related by spin reflection. It is then easy to see that the projected operator
$\mathcal{P}_1Q_4^+\mathcal{P}_1$ is equivalent to \eqref{cXXZ} with $l=1$ and $\Delta=0$. The spin
reflected version obtained from $Q_4^-$ is equivalent to a spin reflected constrained XXZ model. Thus $Q_4$ can be
considered the spin reflection invariant version of \eqref{cXXZ} without any constraints.

The connection between the two models was already noted in \cite{fracton1} and in \cite{folded1,folded2}.
The work \cite{fracton1} actually considered a perturbation of \eqref{H} by an interaction term, such that after
the projection the model becomes equivalent to the constrained XXZ chain with finite $\Delta$.
It was shown in \cite{fracton1} that this full model with the interaction term is not integrable, and only the
constrained sector is Bethe Ansatz solvable. However, it was not understood in  \cite{fracton1} that for the special
point of $\Delta=0$ (corresponding to our $Q_4$) the full model is integrable and Bethe Ansatz solvable; this point was
correctly given in  \cite{folded1,folded2}.

\section{Dynamics and Bethe Ansatz}

\label{sec:dynamics}

Let us investigate the dynamics of the model.  The hopping term $Q_4$ describes spin exchange between neighbouring
sites, and this can be interpreted as particle hopping.

There are two ways to identify fundamental particles in the
model, by choosing two different reference states. We can choose the reference state $\ket{\emptyset}$ consisting of
all spins up, or the state  $\ket{\bar \emptyset}$ with all spins down. Single particle spin wave excitations can be
constructed above either vacuum state.

Focusing on the state $\ket{\emptyset}$ we can define an $N$-particle basis as
\begin{equation}
  \label{xnot}
  \ket{x_1,\dots,x_N}=\sigma^-_{x_1}\dots \sigma^-_{x_N}\ket{\emptyset},
\end{equation}
where we apply the restriction $1\le x_1<\dots < x_N\le L$ to avoid double counting.

\subsection{Single particle states}

Single particle states with lattice momentum $p$ are given simply as
\begin{equation}
  \ket{p}\equiv \sum_x  e^{ipx} \ket{x}.
\end{equation}
The associated energy is found to be
\begin{equation}
  \label{ep}
  E=e(p),\qquad e(p)=-\cos(p).
\end{equation}
It is useful to define a semi-classical (bare) speed, which describes the propagation of wave packets. It is given by
the well known expression
\begin{equation}
  v(p)=\frac{de(p)}{dp}=\sin(p).
\end{equation}
In a finite volume $L$ the quantization condition is simply
\begin{equation}
  e^{ipL}=1.
\end{equation}

\subsection{Two-particle scattering states}

Let us then focus on the scattering of two spin waves. It is clear from the Hamiltonian that as the two incoming particles
approach each other, they can not occupy neighbouring sites. The reason is that any amplitude which would bring two
particles to neighbouring sites is forbidden by the form of $Q_4$. Nevertheless there will be an interaction between the
two particles: the exact wave function for the scattering of particles with momenta $p_1$ and $p_2$ is found to be
\begin{equation}
  \label{2ptwave}
\ket{p_1,p_2}=\sum_{x_1<x_2}\chi(x_1,x_2) \ket{x_1,x_2}
\end{equation}
with
\begin{equation}
  \label{2ptchi}
  \chi(x_1,x_2)=e^{i(p_1x_1+p_2(x_2-1))}-e^{i(p_1(x_2-1)+p_2x_1)}.
\end{equation}
It can be checked that this is an exact eigenstate with the energy given by
\begin{equation}
  E=e(p_1)+e(p_2).
\end{equation}
This is close to a free fermionic wave function, but there are extra phases in the two terms, which can not be
transformed away. In fact we can read off the two-body scattering factor
\begin{equation}
  \label{S11}
  S(p_1,p_2)=e^{i\delta(p_1,p_2)}=-e^{-i(p_1-p_2)}.
\end{equation}
The physical meaning of the scattering phase can be read off directly the wave function \eqref{2ptchi}. If we identify
the first term as an incoming wave and the second term as the outgoing wave (corresponding to $v_1>v_2$), then we see
that both particles suffer a displacement of $\pm 1$ sites. This displacement is most easily understood in a
semi-classical picture, by constructing wave packets and looking at their peaks before and after the scattering. We
observe that the particle coming from the left (right) is moved by 1 site to the right (left). This is consistent with
attractive interactions in a semi-classical or classical picture.

It is worthwhile to recall that generally the scattering displacements can be computed from
the derivatives of the scattering phase $\delta(p_1,p_2)$ with respect to
the momenta \cite{wigner-time-delay}. Generally such a scattering also results in the distortion of the wave packet. In
contrast, now the function $\delta(p_1,p_2)$ is linear in both variables, and we obtain a
complete and exact displacement of the wave packets.

Note that the scattering phase is such that the neighbouring positions $x_2=x_1+1$ are automatically discarded, they
receive zero amplitude. Therefore it is not necessary to impose the constraint by hand as in \eqref{cXXZ}, in this
process  it emerges dynamically, and the resulting $S$-matrix phase is identical to the one of the constrained XXZ model
with $\Delta=0$ and $\ell=1$.

In a finite volume $L$ the periodicity condition for the wave function results in the Bethe Ansatz equations
\begin{equation}
  e^{ip_1L}S(p_1,p_2)=1,\qquad  e^{ip_2L}S(p_2,p_1)=1.
\end{equation}
Using the concrete form of the scattering phase, and denoting $e^{ip_1}e^{ip_2}=e^{iP}$ we get the equivalent set of equations
\begin{equation}
  e^{iPL}=1,\qquad e^{ip_1 (L-2)}=e^{ip_2(L-2)}=-e^{-iP}.
\end{equation}
We see that the total momentum is quantized as usually, but for the single particle momenta the conditions are different
from a free theory. The volume appearing in the conditions is changed by $-2$, and there appears a twist which depends
on the total momentum. These two changes signal that there is indeed true interaction in the model, even though the wave
function appears almost free.

\subsection{$N$-particle scattering states}

The previous wave function can be written as a determinant
\begin{equation}
  \chi(x_1,x_2)=
  \begin{vmatrix}
e^{ip_1 x_1} & e^{ip_1 (x_2-1)} \\
e^{ip_2 x_1} & e^{ip_2 (x_2-1)}
\end{vmatrix}.
\end{equation}
It turns out that this structure can be generalized to higher particle states. The $N$-particle sector is found to be
integrable, with elastic factorized scattering, with the two-body $S$-matrix given by \eqref{S11}. This will be proven
in Section \ref{sec:XXZ}. As a result, the $N$-particle wave function is written as a Vandermonde-like determinant
\begin{equation}
  \label{Bethestate}
  \chi(x_1,\dots,x_N)=\det C,
\end{equation}
where $C$ is a matrix of size $N\times N$ with elements
\begin{equation}
  C_{jk}=e^{ip_j (x_k-k+1)}.
\end{equation}
This is the same wave function as given in
\cite{pronko-abarenkova-ising-limit} and related works.
The energy of such a state is given by
\begin{equation}
  E=\sum_{j=1}^N e(p_j).
\end{equation}
We note again that the wave function is such that neighbouring particle positions $x_{j+1}=x_j+1$ are automatically
forbidden, and this condition naturally emerges from the dynamics of the model.

In finite volume the Bethe equations can be written as
\begin{equation}
  e^{ip_jL} \prod_{k\ne j} S(p_j-p_k)=1.
\end{equation}
Introducing again the total momentum
\begin{equation}
  \label{sump}
  P=\sum_{j=1}^N p_j,
\end{equation}
we get the equivalent set
\begin{equation}
  e^{iPL}=1,\qquad e^{ip_j(L-N)}=(-1)^{N-1} e^{-iP}.
\end{equation}
We see again that the quantization conditions are almost free. Every momentum $p_j$ is quantized with a simple relation
where a modified volume $L-N$ appears, together with a twist depending on the particle number and the overall
momentum. Each quantization condition can then be solved almost independently, with the only requirement that the
overall constraint \eqref{sump} is satisfied.

The appearance of a modified volume signals a relation to the $T\bar T$ or hard rod deformations. We discuss this
connection in more detail in Section \ref{sec:ttbar}.

\subsection{Domain walls}

So far we investigated scattering states consisting of separate particles. In these states neighbouring positions are
forbidden. However, configurations with neighbouring down spins are not excluded from the Hilbert space, thus we need to investigate them
separately.

It turns out that in this model blocks of spins with the same orientation are completely frozen if the block length is
at least 2. The simplest example is probably the case of two down spins on neighbour sites, for example
\begin{equation}
  \label{s12}
  \ket{1,2}=\sigma^-_1\sigma^-_2\ket{\emptyset}.
\end{equation}
Direct calculation shows that this state is an eigenstate of $Q_4$ with eigenvalue 0. The stability of this state
originates from the control on the particle jumps in $Q_4$: neither down spin is allowed to hop, because the control
spins 
result in zero amplitude. Similarly, the same eigenvalue zero is obtained for any state which has $n$ consecutive down
spins embedded in the vacuum state.

To understand the situation we introduce the concept of the Domain Wall (DW): We call a DW the boundary between two
regions of the chain with different spin orientation, such that each region has at least length $2$. With this
definition the state \eqref{s12} can be considered as having two DW's at a distance 2 from each other.
In this picture a single particle can be interpreted as a bound state of two DW's, such that the DW's are at
neighbouring positions. However, it is best to distinguish this situation and call the particle simply a
``particle''. The frozen blocks of spins can be regarded as bound states of particles.

In a finite volume the number of the domain walls is always even, but in infinite volume we can also have
an odd number of them, when the asymptotic reference states on the left and the right are different.

The states of the computational basis with an arbitrary number of DW's at arbitrary positions are eigenstates of $Q_4$
with eigenvalue 0, given that the state does not include any particles. This results in a huge degeneracy of the
reference state, which scales exponentially with the volume (see \cite{fracton1,folded1,folded2} and our discussion in
Sec. \ref{sec:deg}).

In order to fully understand the dynamics of the model we need to consider the scattering of particles with domain
walls. The simplest situation is the case with one particle and one DW. In this situation there is an incoming particle
which meets a standing domain wall. Let us situate the domain wall initially at position 2. Then the incoming wave is
written as
\begin{equation}
\sum_{x=-\infty}^{0} e^{ipx}\ket{x,2,3,4,\dots}.
\end{equation}
Note that the incoming particle is not allowed to hop to the site $x=1$, because this process is forbidden by the
control in the Hamiltonian. However, if the particle occupies $x=0$, then the up spin at site $x=1$ can start to
propagate into the vacuum formed by the down spins on the right. Thus a hole will propagate as an outgoing wave.
We can find the total wave function to be
\begin{equation}
  \label{pDW1}
  \begin{split}
      \ket{\Psi}=&\sum_{x=-\infty}^{0} e^{ipx}\ket{x,2,3,4,\dots}\\
&+  \sum_{x=2}^{\infty}   e^{ip(x-1)}\ket{0,1,\dots,\hat x,\dots},
\end{split}
\end{equation}
where it is now understood that $\hat x$ is missing in the list, that is the position of the hole. Direct computation
shows that this wave function is an exact eigenstate with energy $E=e(p)$.

\begin{figure}[h!]
\centering
\includegraphics[scale=0.22]{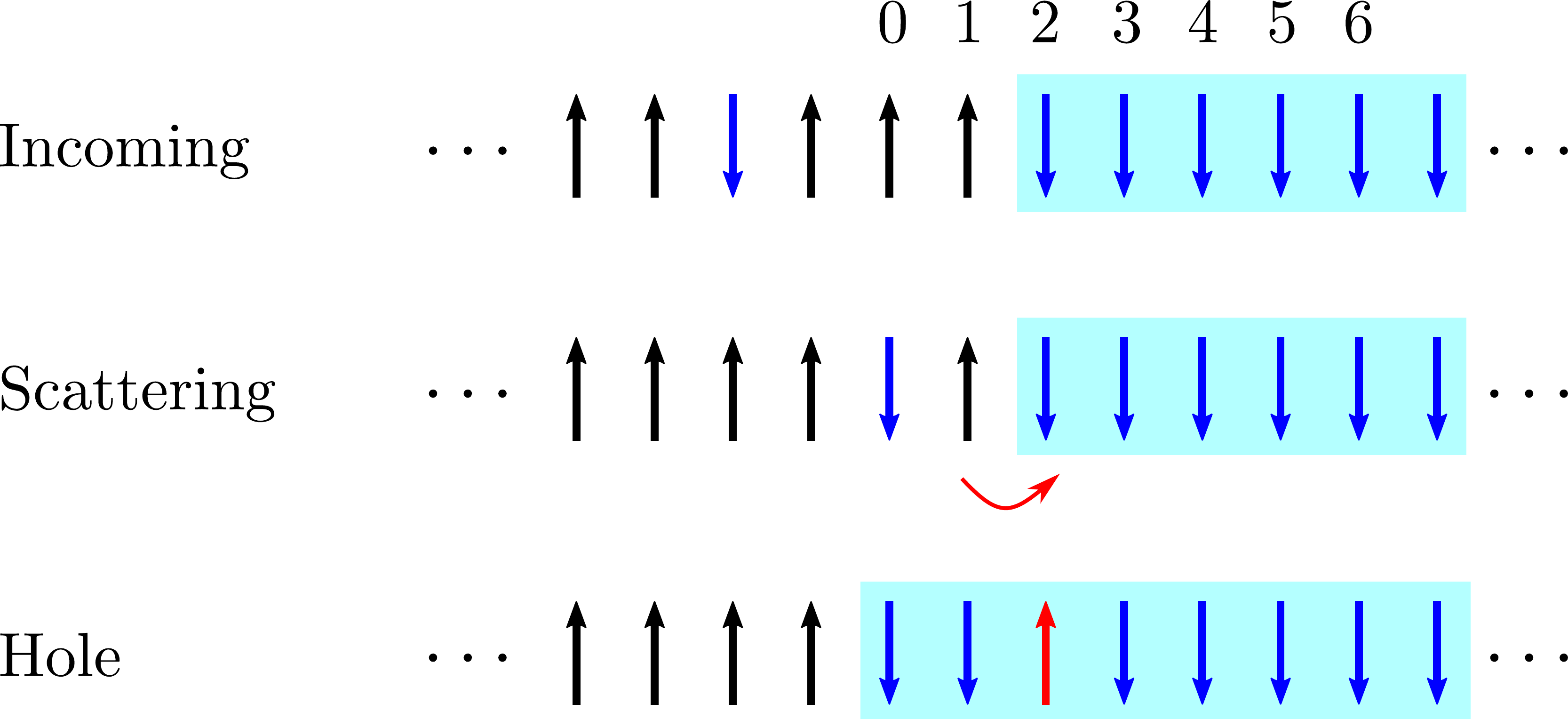}
\caption{Schematic representation of the scattering of a particle with a domain wall. When the particle arrives at position 0, the jump to site 1
  is forbidden by the control factor in \eqref{Q4}. However, in this case the up spin at site 1 becomes mobile and it
  will start to propagate to the right. Eventually the domain wall gets replaced by 2 sites to the left, and the
  trajectory of the particle receives a displacement of 1 site to the right.
}
\label{fig:DW}
\end{figure}

The interpretation of this wave function is the following: As a result of the scattering the domain wall gets displaced
by 2 sites, and the particle also obtains a displacement of 1 site forwards, see figure~\ref{fig:DW}.
Interestingly, the domain wall does not contribute to the energy, but it has an effect on the dynamics due to the
scattering displacement. The scattering event is shown schematically in Fig. \ref{fig:DW}.

%\red{An example of the scattering of a single particle and a length-4 bound state is given in figure~\ref{fig:boundstate2}.}
%\begin{figure}[h!]
%\begin{center}
%\includegraphics[scale=0.22]{boundstate2.pdf}
%\caption{The scattering of a single particle with a length-4 bound state. There are configurations which correspond to the propagation of holes}
%\label{fig:boundstate2}
%\end{center}
%\end{figure}

\begin{figure}[h!]
\begin{center}
\includegraphics[scale=0.22]{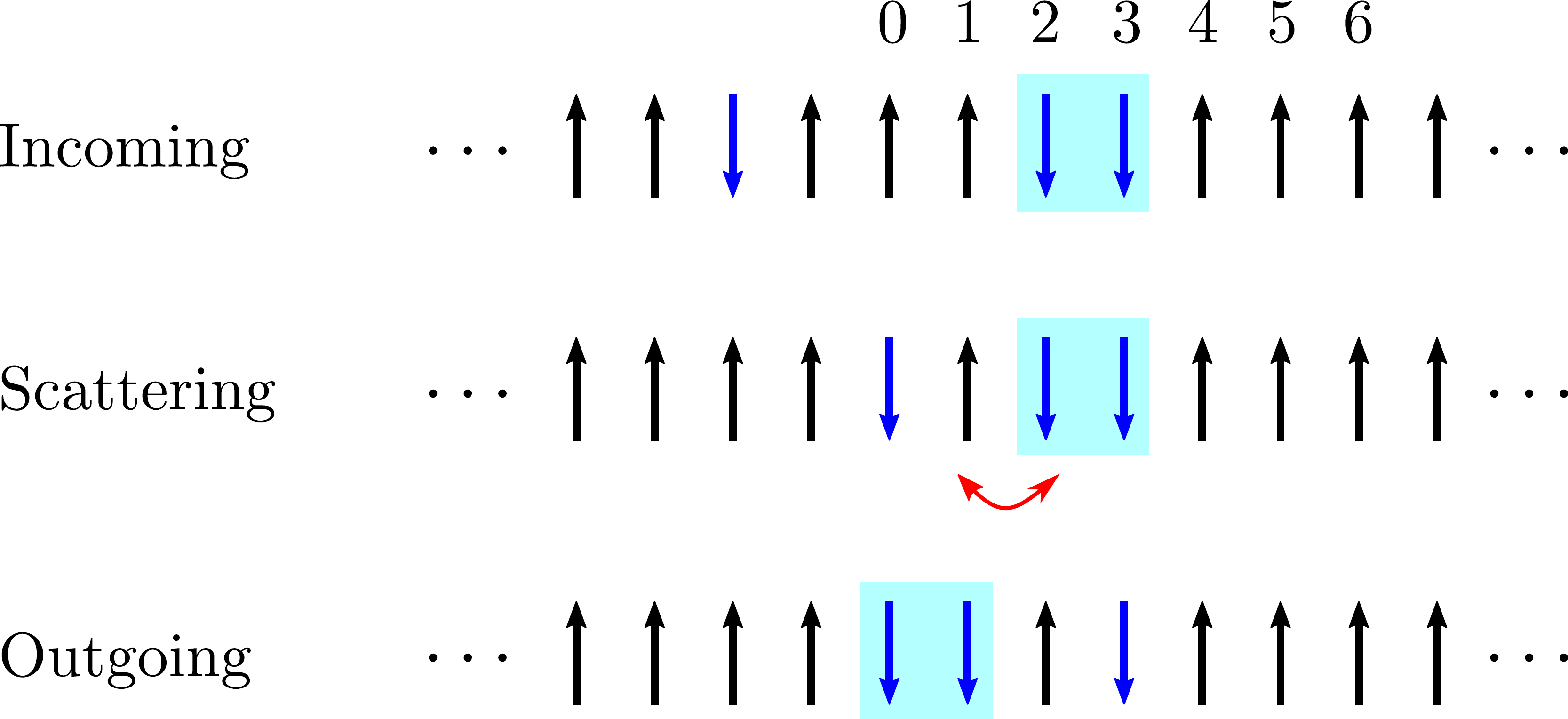}
\caption{Scattering of a single particle and a length-2 bound state.}
\label{fig:boundstate}
\end{center}
\end{figure}

We can also compute the situation with one particle and two domain walls, which is equivalent to having a particle and a
bound state with a given number of down
spins. For example if the bound state is of length two, we obtain the exact wave function
\begin{equation}
  \label{pDW2}
  \ket{\Psi}=\sum_{x=-\infty}^{0} e^{ipx}\ket{x,2,3}+  \sum_{x=3}^{\infty}   e^{ip(x-2)}\ket{0,1,x}.
\end{equation}
The interpretation of the wave
function is the following: originally the bound state of two DW's occupies positions 2 and 3, and there is an incoming
wave. Afterwards the bound state occupies positions 0 and 1 and the outgoing wave also suffers a displacement of 2 in
the forward direction. This is shown in figure~\ref{fig:boundstate}.
Notice that the hard core property is still
satisfied, although it is not enforced, it is completely dynamical even in this case. This state also has energy
$E=e(p)$, the two domain walls do not contribute.

The states \eqref{pDW1} and \eqref{pDW2} only exists in infinite volume, because the periodicity conditions can not be
satisfied by them. However, the state \eqref{pDW2} can be modified to fit into a finite volume situation with periodic
boundaries. To this order
we introduce a Fourier transform over the position of the block of 2 spins and write
\begin{multline}
  \label{pDW3}
  \ket{\Psi}=\sum_{y>x+1} e^{i(px+k(y-2))}\ket{x,y,y+1}\\
  +\sum_{x>y+2}   e^{i(p(x-2)+ky)}\ket{y,y+1,x}.
\end{multline}
We can extract from this wave function the scattering phase
\begin{equation}
  \label{2DW}
   S_{p,2\text{DW}}(p,k)=e^{-2i(p-k)}.
\end{equation}
The state \eqref{pDW3} has energy $E=e(p)$ and it is periodic if $p$ and $k$ satisfy the equations
\begin{equation}
  e^{ipL}S_{p,2\text{DW}}(p,k)=1,\qquad  e^{ikL}S_{p,2\text{DW}}(k,p)=1.
\end{equation}
Substituting \eqref{2DW} into the quantization conditions and introducing once again the total momentum $P=p+k$ we obtain
\begin{equation}
  e^{ip(L-3)}=e^{-iP},\qquad   e^{ik(L-3)}=e^{-iP}.
\end{equation}
The domain walls do not contribute to the energy, but they modify the propagation of the free particles, and
also the quantization conditions.
The effect of the interaction is a change
in the apparent length of the system by -3 and the appearance of the total momentum as a twist for the quantization conditions.

It turns out that this picture can be generalized to many body states: the model is
integrable, and scattering of the particles among each other and on the domain walls leads to the same phases that we
computed in this Section. In particular the scattering phase between a particle and a frozen block of spins of length
$n$ does not depend on $n$.

The complete integrability of the model is most easily understood by considering a special
connection to the XXZ spin chain, which we discuss in the next Section. The degeneracies resulting from the presence of
the domain walls in generic states is discussed later in Section \ref{sec:deg}.

\section{Relation to the XXZ spin chain}

\label{sec:XXZ}

Here we show that our model can be derived as a special limit of the XXZ spin
chain. Our procedure is different from the one in \cite{folded1,folded2}, even though we also consider the large
anisotropy limit.

The XXZ chain is given by the Hamiltonian
\begin{equation}
  \label{HXXZ}
  H=\sum_j \sigma^x_j\sigma^x_{j+1}+\sigma^y_j\sigma^y_{j+1}+\Delta (\sigma^z_j\sigma^z_{j+1}-1),
\end{equation}
where $\Delta$ is the anisotropy parameter.

It is known that the XXZ chain is integrable, and it possesses a set of conserved charges that will be denoted as
$\tilde Q_\alpha$. They  satisfy
\begin{equation}
  \label{XXZcomm}
  [\tilde Q_\alpha,\tilde Q_\beta]=0.
\end{equation}
The charge $\tilde Q_\alpha$ has an operator
density that spans $\alpha$ sites; as usually $\tilde Q_2$ can be identified with the Hamiltonian, and $\tilde Q_1$ can
be chosen as the global $S^z$ or alternatively as $Q_1$ written in \eqref{charges12}.

The charges $\tilde Q_\alpha$ can be obtained either from a transfer matrix, or using the so-called boost operator
\cite{Tetelman,Thacker-boost,GM-higher-conserved-XXZ}; we describe here the latter method. The boost operator is defined
as the formal expression
\begin{equation}
  \mathcal{B}=i\sum_j j \tilde h_{j,j+1},
\end{equation}
where $\tilde h_{j,j+1}$ is the Hamiltonian density acting on sites $j,j+1$. Then the charges are constructed through
the recursive relations
\begin{equation}
  \label{boostrel}
  \tilde Q_{\alpha+1}=[\mathcal{B},\tilde Q_\alpha].
\end{equation}
Even though the r.h.s. is just a formal expression, the actual commutation relations result in a well defined extensive
and local charge at each new step. This relation and the resulting charges were studied in detail in
\cite{GM-higher-conserved-XXX,GM-higher-conserved-XXZ}, where concrete formulas were given in a number of different
cases. Furthermore, explicit formulas for $Q_\alpha$ with arbitrary $\alpha$ were derived in the recent works
\cite{xyz-all-charges,bernard-charges}.

We obtain our model by performing a special $\Delta\to\infty$ limit on the set of the conserved charges $\tilde
Q_\alpha$.
It can be seen from the boost relation that in the XXZ model the operators $\tilde Q_\alpha$ are polynomials
in $\Delta$.
Our main idea is to select the leading terms in $\Delta$ using a recursive procedure. Then the  commutativity
\eqref{XXZcomm} will then ensure that our new charges $Q_\alpha$ also commute.

We start with $Q_1$
which does not depend $\Delta$, thus it will stay constant during the limit. The next charge is $\tilde Q_2$, which is the Hamiltonian
\eqref{HXXZ}. We select the leading piece in $\Delta$, which is (apart from the additive normalization, and stripping
away the factor of $\Delta$) equal to $Q_2$ as given in
\eqref{charges12}. It is clear that $[Q_1,Q_2]=0$. The charge $Q_2$ is not dynamical: the kinematical piece of the
original $\tilde Q_2$ is scaled to zero, and only the ``classical'' part remains, which describes the classical Ising
model. The disappearance of the dynamical terms in $Q_2$ prompts us to turn to the higher charges of the XXZ
chain.

Therefore we take the explicit representation of
$\tilde Q_3$ and $\tilde Q_4$ found in \cite{GM-higher-conserved-XXZ,xyz-all-charges,bernard-charges} and we select the leading terms in $\Delta$. It
turns out that in the normalization of \cite{GM-higher-conserved-XXZ} the maximum power of $\Delta$ in $\tilde Q_3$ is
linear, whereas it is quadratic in $\tilde Q_4$. Selecting these terms and stripping away the factors of $\Delta$ we
obtain $Q_4$ and a new charge
\begin{equation}
  \begin{split}
     Q_3&=\sum_j
%   (\sigma^z_j+\sigma^z_{j+3})(\sigma^x_{j+1}\sigma^y_{j+2}-\sigma^y_{j+1}\sigma^x_{j+2})
\frac{i}{4} (\sigma^z_j+\sigma^z_{j+3})(\sigma^+_{j+1}\sigma^-_{j+2}-\sigma^-_{j+1}\sigma^+_{j+2}).
   \end{split}
  \end{equation}
   It is clear from the construction that all four charges  commute
with each
other, because their commutativity is simply the leading order term from the relation \eqref{XXZcomm}. And $Q_3$ and
$Q_4$ are already dynamical, even after the $\Delta\to\infty$ limit has been taken. The numerical pre-factors in $Q_3$
and $Q_4$ are added only for convenience, such that they have simple one-particle eigenvalues. Note that $Q_3$ is
actually a three-site operator, it is just written conveniently in a four site representation as above.

Continuing to even higher charges we observe that more steps are needed. For example we find that the leading term in
$\tilde Q_5$ is of order $\Delta^2$, and it is proportional to $Q_3$ as given above. Therefore a new charge can be obtained
only if we subtract from $\tilde Q_5$ the appropriate multiple of $\tilde Q_3$ and then consider the leading term of the
remainder. This procedure leads to a new operator $Q_5$ which commutes with all previous charges and which is given
(after stripping away an irrelevant numerical pre-factor) as
\begin{equation}
\begin{split}
  Q_5=&\sum_j
  (\sigma^z_j+\sigma^z_{j+4})\Big(\sigma^x_{j+1}\sigma^y_{j+3}-\sigma^y_{j+1}\sigma^x_{j+3}\Big)\\
  &+(\sigma^z_j+\sigma^z_{j+5})\Big(\sigma^x_{j+1}\sigma^x_{j+2}+\sigma^y_{j+1}\sigma^y_{j+2}\Big) \\
  &\times \Big(\sigma^y_{j+3}\sigma^x_{j+4}-\sigma^x_{j+3}\sigma^y_{j+4}\Big)\\
&+(1+\sigma^z_{j}\sigma^z_{j+4})\sigma^z_{j+2}\Big(\sigma^x_{j+1}\sigma^y_{j+3}-\sigma^y_{j+1}\sigma^x_{j+3}\Big).
\end{split}
\end{equation}
%The operator density of $Q_5$ appears to span 6 sites ranging from $j$ to $j+5$, but
$Q_5$ appears to be a 6 site operator, but this is only apparent: after expanding the products it can be seen that every single term spans a maximum of 5 sites
only.  It follows from the construction that $Q_5$ commutes with all previous charges of our model. The charge $Q_5$ is
identical to the operator $Q_2^-$ of \cite{folded1}, see formula (31) in that paper.

We can continue this recursive procedure to obtain the higher charges $Q_\alpha$ with $\alpha\ge 6$. The idea is always
to focus on the leading terms in $\Delta$, and to subtract contributions that appeared in the earlier charges. We
conjecture that this procedure leads to an infinite set of linearly independent charges $Q_\alpha$.

At present we do not have
closed form results for the higher charges, and we do not have a direct transfer matrix construction for the new charges
either. However, the existence of the charges $Q_\alpha$ with $\alpha=1\ldots 5$ is already enough to claim the integrability
of the model. In fact, for integrability it is enough to have 2 independent dynamical conserved charges
\cite{kulish-s-matrix}, which in the
present case are $Q_3$ and $Q_4$. The results of the works \cite{xyz-all-charges,bernard-charges} could be used to give explicit
formulas for all the charges of our model, but this is beyond the scope of the present work.

We note a curious property of our construction: the boost relation \eqref{boostrel} does not survive the
$\Delta\to\infty$ limit. This is most easily seen from the relation between $Q_2$ and $Q_3$: whereas the original
$\tilde Q_3$ is obtained from $\tilde Q_2$ using the boost, this is not true after the scaling. The new charge $Q_2$ is
not dynamical, and it does not give $Q_3$ using the relation \eqref{boostrel}. However, we observe that the next
relation is kept intact, which means that
\begin{equation}
  Q_4\sim [\mathcal{B},Q_3]
\end{equation}
with $\mathcal{B}$ given by the boosted $Q_2$.
We comment further on this issue in the Conclusions.

%The first few charges of the model (including $Q_5$ and $Q_6$ in our notations) were computed in \cite{folded1,folded2}.
%There they were found by directly evaluating commutation relations with $Q_4$.
%The direct relation
%with the charges of the XXZ model was not given there, although the connection with the transfer matrix of the 4-vertex
%model (which is the $\Delta\to\infty$ limit of the transfer matrix of the XXZ model) was noted.

\subsection{Bethe Ansatz for the XXZ chain}

Here we present a summary of the Bethe Ansatz solution of the XXZ chain, which is well known
\cite{Takahashi-Book}. Our goal is to consider the $\Delta\to\infty$ limit of the full solution, this is presented
below. Throughout this Section we will use the conventional parametrization
\begin{equation}
  \Delta=\cosh(\eta).
\end{equation}

It is known that the $N$-particle eigenstates of the XXZ model can be
characterized by Bethe rapidities $\lambda_j$, $j=1\dots N$, that describe the one-particle momenta within the interacting states. The spin waves
can form bound states, which are described by the so-called string solutions. The string hypothesis states that in the
thermodynamic limit almost all eigenstates can be described by Bethe roots that organize themselves into the string
patterns.

In our normalization $\lambda\in [-\pi/2,\pi/2]$ and the $n$-string  propagation factors are
\begin{equation}
  e^{ip_n(\lambda)}=\frac{1}{\Theta_{n/2}(\lambda)},
\end{equation}
where we defined
\begin{equation}
  \Theta_n(\lambda)=\frac{\sin(\lambda+in\eta)}{\sin(\lambda-in\eta)}.
\end{equation}
The scattering of an $n$-string and an $m$-string is described by
\begin{equation}
  \begin{split}
&  S_{n,m}=
   \begin{cases}
        \Theta_{\abs{n-m}} \Theta^2_{\abs{n-m}+2} \dots  %\Theta^2_{n+m-2}
     \Theta_{n+m}
    &\text{ for } n\ne m \\
    \Theta^2_{2} \dots \Theta^2_{2n-2} \Theta_{2n} &\text{ for } n=m.
    \end{cases}
 \end{split}
  \end{equation}
Here we omitted the dependence on $\lambda$ in the notations.

We denote the string centers for the $n$-strings as $\lambda_{j,n}$. Then the approximate Bethe equations for the
strings are
\begin{equation}
\label{generalBAxxz}
e^{ip_n(\lambda_{j,n})L} \mathop{\prod_{m,k}}_{(j,n)\ne (k,m)} S_{n,m}(\lambda_{j,n}-\lambda_{k,m})=1.
\end{equation}

The energy of such an eigenstate is
\begin{equation}
  \label{Eh2}
  \tilde E=2\sinh(\eta) \sum_{j,n} \tilde h_{2,n}(\lambda_{j,n})
\end{equation}
where
\begin{equation}
   \tilde h_{2,n}(\lambda)=\coth(-n\eta/2+i\lambda)-\coth(n\eta/2+i\lambda).
 \end{equation}
In a normalization dictated by the boost operator construction (see for example Section 3 of \cite{sajat-xxz-gge}) the eigenvalues
of $\tilde Q_4$ are
\begin{equation}
  \label{Q4b}
  \tilde Q_4=(2\sinh(\eta))^3 \sum_{j,n} \tilde h_{4,n}(\lambda_{j,n})
\end{equation}
where
\begin{equation}
   \tilde h_{4,n}(\lambda)=-\frac{\partial^2 \tilde h_{2,n}(\lambda)}{\partial \lambda^2}
\end{equation}
or with a more explicit formula
\begin{multline}
      \tilde h_{4,n}(\lambda)=
      2\Big[\coth(n\eta/2+i\lambda)- \coth^3(n\eta/2+i\lambda)\\
      -\coth(-n\eta/2+i\lambda)+\coth^3(-n\eta/2+i\lambda)\Big].
\end{multline}

\subsection{The special limit of the Bethe Ansatz}

Here we consider the   $\Delta\to\infty$ limit of the previous formulas.
This particular limit of the XXZ chain was already studied in
a series of works \cite{bog-ising-limit,bog-ising-limit2}. Also, it was known that in this
limit the constrained XXZ chain can also be obtained \cite{pronko-abarenkova-ising-limit,xxz-triple-point}.  However,
the discussion of the dynamics under $Q_4$ is new.

First let us discuss the interpretation of the strings. It is known that in the $\Delta\to\infty$ limit the bound states
become more and more bound
\cite{js-jorn-domainwall}. Eventually in the strict  $\Delta\to\infty$ limit an $n$-string will describe $n$ down spins
placed beside each other. Thus it is natural in this special limit that the string solutions will correspond to the
bound states in our model. Now we compute this correspondence, and the associated scattering phases.

We put forward that one limitation of this Bethe Ansatz picture is that it does not mirror the spin reflection symmetry
of our model. As a consequence, the simple scattering state \eqref{pDW1} of a particle and a DW is a very complicated object
in this picture: the sea of down spins on the right would be described by an infinitely large string.  In Section
\ref{sec:bond} we present an alternative representation of the same model, where a single DW is treated as a separate
particle. This picture will avoid some of the complications of the present description.

Now we compute the special limit of the eigenvalue functions. First of all we find that
\begin{equation}
\lim_{\Delta\to\infty} \tilde  h_{2,n}(\lambda)= -2.
\end{equation}
The explicit form of $Q_2$ given in \eqref{charges12} implies the limit
\begin{equation}
  Q_2=\lim_{\Delta\to\infty} -\frac{H_{XXZ}}{2\Delta}
\end{equation}
which together with \eqref{Eh2} gives
\begin{equation}
  h_{2,n}(\lambda)=2.
\end{equation}
We see that the particles and the bound states give an equal contribution to $Q_2$, which does not depend on the
momentum. This is clear from the actual form of $Q_2$, which shows sensitivity only to the domain walls.

Regarding the scaling of the $\tilde Q_4$ eigenvalues we find the large $\eta$ behaviour
\begin{equation}
 \tilde  h_{4,n}(\lambda)\approx -16e^{-n\eta} \cos(p),
\end{equation}
where we identified $2\lambda=p$.

Combining with \eqref{Q4b} and a re-scaling of $\tilde Q_4$ according to
\begin{equation}
  Q_4=\lim_{\Delta\to\infty} \frac{\tilde Q_4}{64 \Delta^2}
\end{equation}
we get the scaling
\begin{equation}
  h_{4,n}(p)=\lim_{\Delta\to\infty} \frac{(2\sinh(\eta))^3\tilde h_{4,n}(\lambda)}{64\Delta^2}
\end{equation}
leading to
\begin{equation}
 h_{4,n}(\lambda)=
  \begin{cases}
    -\cos(p)&\text{ for } n=1\\
     0 &\text{ for } n>1.
  \end{cases}
\end{equation}
This coincides with our real space computations in Section \ref{sec:dynamics}.

Let us now consider the Bethe equations. Note that the limit of the functions
$\Theta_n$ is simply
\begin{equation}
  \Theta_n(\lambda)\quad\to\quad -e^{-ip},
\end{equation}
where we identified $p=2\lambda$.

For the limit of the $S$-matrix factor between 1-strings we find simply:
\begin{equation}
  S_{1,1}(p)=-e^{-ip}.
\end{equation}
This agrees with the scattering phase found earlier in \eqref{S11}.

For the limit of the $S$-matrix factor between 1-strings and $n$-strings for $n>1$ we find:
\begin{equation}
  \label{S1n}
  S_{1,n}(p)=e^{-2ip}.
\end{equation}
This is the same factor as found earlier in \eqref{2DW}. We stress that this factor does not depend on $n$: the
scattering phase is independent of the length of the bound state, it only depends on the number of domain walls crossed,
which is two in this case.

We  also compute the
$S$-matrix between an $n$-string and an $m$-string with $n,m>1$. We find
\begin{equation}
  S_{m,n}=
  \begin{cases}
    e^{-2m ip} &\text{ for } n>m\\
    -e^{-2nip} & \text{ for } n=m.\\
  \end{cases}
\end{equation}
Then we obtain the final Bethe equations
\begin{equation}
\label{generalBA}
e^{ip_{j,n}L} \mathop{\prod_{m,k}}_{(j,n)\ne (k,m)} S_{n,m}(p_{j,n}-p_{k,m})=1.
\end{equation}
For particles (or equivalently for 1-strings) the Bethe equations take a particularly simple form. Using the above
results we get
\begin{equation}
  e^{ip_{j,1}L} \prod_{k\ne j} -e^{-i(p_{j,1}-p_{k,1})} \prod_{m=2,3,\dots}\prod_{k} e^{-2i(p_{j,1}-p_{k,m})}=1.
\end{equation}
This is then rewritten as
\begin{equation}
  e^{ip_{j,1}(L-N_1-2N_s)}=(-1)^{N_1-1} e^{-iP_1}e^{-2iP_{s}},
\end{equation}
where we introduced the total string number (without the particles)
\begin{equation}
  N_s=N_2+N_3+\dots
\end{equation}
and
\begin{equation}
  P_1=\sum_j p_{j,1},\qquad P_s=\sum_{m=2,3,\dots}\sum_k p_{k,m}.
\end{equation}
We see that for the quantization condition of these particles the various strings are grouped together, such that the
information on the string length disappears. Only the total number of the strings and their total momentum enters the
equations. This is perfect agreement with other descriptions of the model which will be discussed below.

The energy of the states is
\begin{multline}
  \label{Egen}
  E=\sum_{j} e(p_{j,1})+\mu (N_1+N_s)\\ +h \left[N_1+\sum_{m=2,3,\dots} m N_m \right].
\end{multline}
We observe that degeneracies between different string lengths remain if $h=0$. This is discussed in more detail below in
\ref{sec:deg}.

\section{Boundary conditions}

\label{sec:boundary}

In this Section we investigate the model with open boundaries; this is the case that was originally considered in
\cite{fracton1}. We will show that the boundary problem is even simpler than the periodic one: it gives simpler Bethe
equations, we can easily prove completeness of the Bethe Ansatz, and the study of the thermodynamics is also easier (see
Section \ref{sec:thermo}).

This simplicity can be explained by a standard semi-classical picture. We consider the motion of the
particles in a finite volume, both in the periodic and in the boundary case. In the periodic case the particles move
along the chain and eventually travel around the full volume. In this process every particle hits every DW one time,
therefore the domain walls are displaced. This process is then repeated over time, thus the DW's can not have fixed
positions. In contrast, in the boundary case the particles travel back and forth between the left and right boundaries.
In this process every particle hits every DW one time from the left and one time from the right, and the displacements
of
the DW's add up to zero. In other words the DW's stay in their
original positions as the particles finish a full cycle of traversing the available volume. This means that in the
boundary case we can label
the states with the DW positions and the particle momenta.
This phenomenon shows that the open boundary condition is ''more compatible'' with the bulk scattering
than the periodic one.

In the boundary setting the Hamiltonian of a chain of length $L+2$ is
\begin{equation}
H =-\frac{1}{4}\sum_{j=0}^{L-1}\left(1+\sigma_{j}^{z}\sigma_{j+3}^{z}\right)\left(\sigma_{j+1}^{+}\sigma_{j+2}^{-}+\sigma_{j+1}^{-}\sigma_{j+2}^{+}\right).
\end{equation}
This Hamiltonian can be obtained from a double row transfer matrix (using the identity as the so-called $K$-matrices) of the
XXZ spin chain in the same way as for the periodic case (see section \ref{sec:XXZ}). Using the same argument as section
\ref{sec:XXZ} we can convince ourselves that the Hamiltonian is integrable. We will not treat this procedure in detail
here, and for the algebraic framework of the boundary
XXZ chain we refer the reader to \cite{sklyanin-boundary}.

The following charges are also conserved for open boundaries
\begin{equation}
Q_1=\sum_{j=0}^{L+1} \sigma_j^z,\qquad
  Q_{2} =\sum_{j=0}^{L}\sigma_{j}^{z}\sigma_{j+1}^{z}.
\end{equation}
In addition, specific to the boundary case, we have the following conserved charges
\begin{equation}
B_{l}=\sigma_{0}^{z},\qquad B_{r}=\sigma_{L+1}^{z}.
\end{equation}
We see that the first and the last sites are not dynamical and the
Hilbert space is decomposed as  $\mathcal{H}=\mathcal{H}_{\uparrow\uparrow}\otimes\mathcal{H}_{\uparrow\downarrow}\otimes\mathcal{H}_{\downarrow\uparrow}\otimes\mathcal{H}_{\downarrow\downarrow}$
where
\begin{align}
\mathcal{H}_{\uparrow\uparrow} & =\mathrm{span}\left\{ \left|\uparrow,\dots,\uparrow\right\rangle \right\} , 
& \mathcal{H}_{\uparrow\downarrow} & =\mathrm{span}\left\{ \left|\uparrow,\dots,\downarrow\right\rangle \right\} \nonumber,\\
\mathcal{H}_{\downarrow\uparrow} & =\mathrm{span}\left\{ \left|\downarrow,\dots,\uparrow\right\rangle \right\} ,
& \mathcal{H}_{\downarrow\downarrow} & =\mathrm{span}\left\{ \left|\downarrow,\dots,\downarrow\right\rangle \right\} .
\end{align}
Each subspace has dimension $2^L$.
In the following we focus on the subspace $\mathcal{H}_{\uparrow\uparrow}$. The other sectors can be analyzed similarly.

\subsection{Bethe Ansatz}

In this subsection we present the characterization of the spectrum of $\mathcal{H}_{\uparrow\uparrow}$.
Let us start with the one-magnon state. The Bethe Ansatz eigenstate reads
\begin{equation}
\left|\Psi\right\rangle =\sum_{x=1}^{L}\left(e^{-ipx}+R(p)e^{ipx}\right)\left|x\right\rangle.
\end{equation}
We can calculate the energy, the reflection factor $R(p)$ and the Bethe Ansatz equation in the usual way, which leads to
\begin{align}
 e(p) &= -\cos(p),\\
 R(p) &= -1, \\
 e^{-2ip(L+1)} &= 1.
\end{align}
Solutions to be Bethe equation are
\begin{equation}
 p=\pi\frac{n}{L+1},\qquad n=1,\dots,L.
\end{equation}
The number of solutions agrees with the dimension of the one-magnon subspace of $\mathcal{H}_{\uparrow\uparrow}$.

We can generalize this Ansatz to general states.
We learned from the periodic case that the spectrum contains magnons and DW's.
The width of the DW's have to be at least two since the one distance means there
is a hole excitation which will be transformed to a magnon at some point.

There is now a key difference as opposed to the periodic case; above we have already touched upon this issue when
discussing the semi-classical picture. In the boundary setting we can have bound states with fixed
positions, up to the scattering displacements with the particles. In contrast, in the periodic case we needed a Fourier
transform over the bound state positions such as in \eqref{pDW3}, otherwise there was no way to satisfy the periodicity
conditions. At the heart of this problem lies the periodic property itself, which makes it impossible to identify which
particle is to the left or to the right as compared to any other particle. In the boundary setting this is clearly
possible, and we can assign well defined positions to the DW's, which are changed only as a result of the scattering
with the particles.

Now we present the construction of the Bethe states. We present the final results without proofs; the form of the wave
functions follow from the previous results given above.
A long and detailed proof would not contribute considerably to the understanding of the model. However, to
avoid simple mistakes we checked the wave functions below in small volumes and found that they indeed produce the
eigenstates of the boundary model.

The wave functions of the $N$ particle states without DW's are
\begin{equation} \label{eq:opencoba}
|\mathbf{p}_{N}\rangle=\sum_{x_{1}<\dots x_{N}}
\chi(\mathbf{x},\mathbf{p})
|x_{1},x_{2}+1,x_{3}+2,\dots x_{N}+N-1\rangle
\end{equation}
where
\begin{equation}
\chi(\mathbf{x},\mathbf{p}) =
\sum_{\sigma\in\mathcal{S}_{N}}\sum_{s_{j}=\{-,+\}}\mathrm{sgn}(\sigma)\left[\prod_{j=1}^{N}s_{j}\right]
e^{i\sum_{i=1}^{N}s_{i}p_{\sigma(i)}x_{i}}.
\end{equation}
The summation over the signs $s_j$ correspond to whether the particle is moving from the left boundary to the right one
or vice versa.

Going further we can also insert bound states into the chain, which will be displaced as particles scatter on them. Such
states will thus be given by a complicated summation over the various possibilities for the positions of the
particles. In any case the states can be labeled by the positions of the DW's as the particles occupy the leftmost
possible positions. Thus we can label the states with domain walls at positions $(x_{i},y_{i})$ for $i=1,\dots,M$ as
\begin{multline}
|\mathbf{p}_{N},\mathbf{x}_{M},\mathbf{y}_{M}\rangle \approx \\
|1,3,\dots,2N-1,x_{1},x_{1}+1,\dots,y_{1},\dots,x_{M},\dots y_{M}\rangle+\dots
\end{multline}
where $y_{i}-x_{i}\geq2$, $x_{i+1}-y_{i}\geq2$, and the dots stand for other components of the wave function.

\begin{figure}[h!]
\begin{center}
\includegraphics[width=0.85\columnwidth]{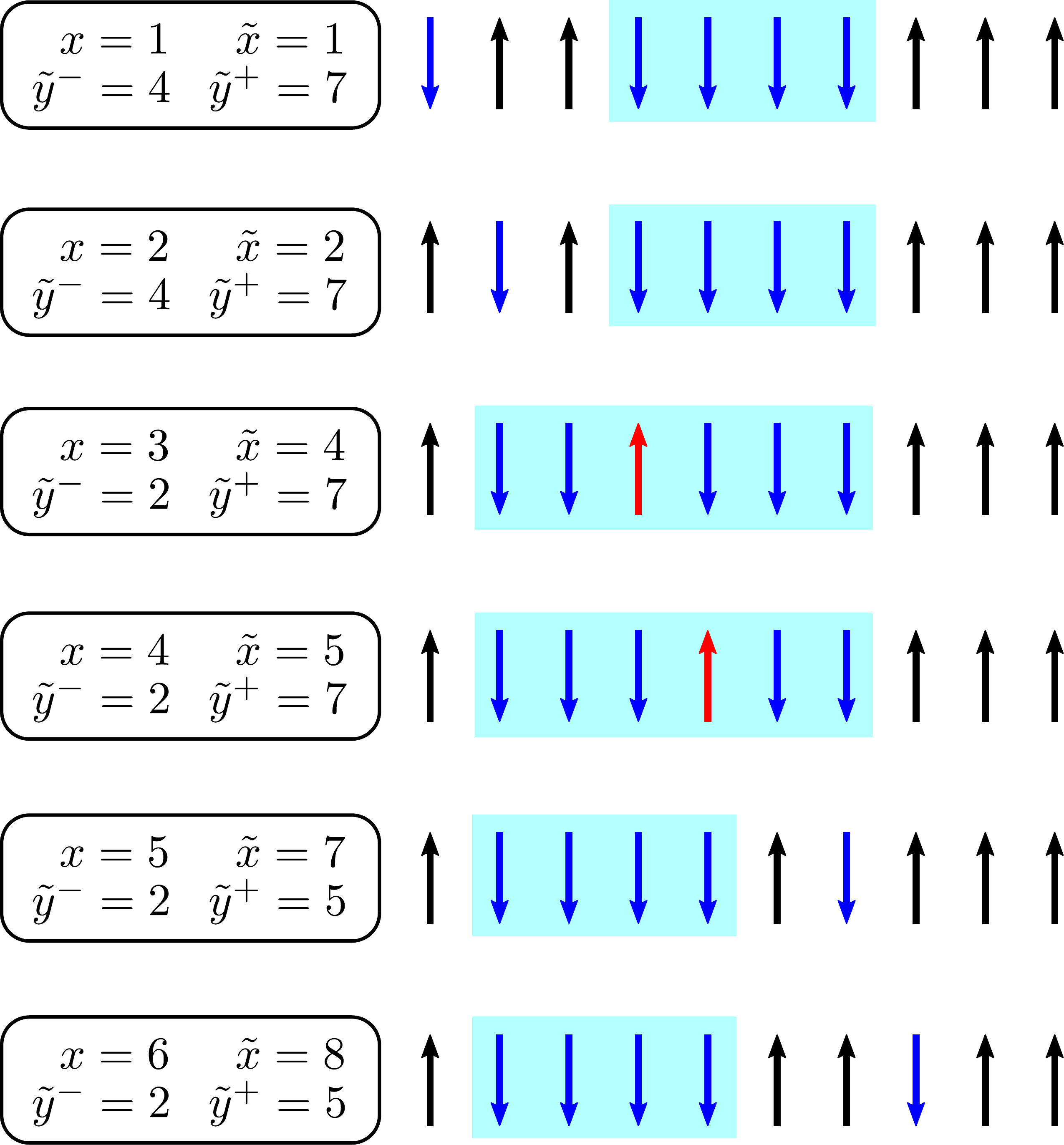}
\caption{Illustration of effective position $x$ and real positions $\tilde{x},\tilde{y}^\pm$.}
\label{fig:effpos}
\end{center}
\end{figure}

To describe the full wave function with $N$ particles and $M$ DW's it is best to consider the motion of $N$ free
fermions in a box with effective length $\tilde{L}=L+1-N-2M$. It is convenient to use effective positions $1\leq
x_1<\dots<x_N\leq \tilde{L}$ and real positions $1\leq\tilde{x}_1<\dots<\tilde{x}_N\leq L$. The effective position is
the position of the particle in the auxiliary free fermion picture and the real position is its actual
position. Ignoring the DWs
we saw that the connection between real and effective positions is $\tilde{x}_k=x+k-1$ (see \eqref{eq:opencoba}) which
is a manifestation of the hard rod property (see also Section \ref{sec:ttbar}).

We also saw that the DWs are displaced by the scattering. Let $(\tilde{y}_j^+,\tilde{y}_j^+)$ be the position of the DW
when the effective positions are $\{x_k\}$. Figure \ref{fig:effpos} shows the connections between these positions if
we have one particle and one bound state. We can see that the real position is increased by one when the particle
reaches the
left of the domain wall and one more when it reaches the right. We can also see that $\tilde{y}^-$ and $\tilde{y}^+$ are
decreased by two when the particle reaches the left and right respectively.

Having more particles and bound states we have to count how many particles reached a DW to get the real positions of
given
effective positions. If the $k$th particle reaches the left of $j$th DW then the $k+1,\dots,N$ particles have to already
reached it therefore its position when the $k$th particle reaches it is $\tilde{y}^-_l=y^-_l-2(N-k)$. The $k$th particle
can reach left of $j$th DW only when it already went through $1,\dots,j-1$ bound states therefore its real position is
$\tilde{x}_k=x_k+k-1+2(j-1)$ therefore the $k$th particle pasted the left of the $j$th DW if $x_{k}-y_{j}^{-}+2N-k+2j-2
\geq 0$. In an analogous way the $k$th particle is pasted the right of the $j$th DW if $x_k-y_{j}^{+}+2N-k+2j-1 \geq
0$.

In summary the full wave function can be written as
\begin{equation}
|\mathbf{p}_{N},\mathbf{y}_{M}^{-},\mathbf{y}_{M}^{+}\rangle =
\sum_{1\leq x_{1}<\dots x_{N}\leq\tilde{L}}
\chi(\mathbf{x},\mathbf{p})
|\tilde{\mathbf{x}}_{N},\tilde{\mathbf{y}}_{M}^{-},\tilde{\mathbf{y}}_{M}^{+}\rangle,
\end{equation}
where $\tilde{L}=L+1-N-2M$, $y_{i}^{+}-y_{i}^{-}\geq2$, $y_{i+1}^{-}-y_{i}^{+}\geq2$,
\begin{align}
\tilde{x}_{k} & =x_{k}+k-1+\sum_{j=1}^{M}\Bigl(\Theta(x_{k}-y_{j}^{-}+2N-k+2j-2), \nonumber\\
&\qquad\qquad\qquad+\Theta(x_k-y_{j}^{+}+2N-k+2j-1)\Bigr),\\
\tilde{y}_{j}^{-} & =y_{j}^{-}-2\sum_{k=1}^{N}\Theta(x_{k}-y_{j}^{-}+2N-k+2j-2),\\
\tilde{y}_{j}^{+} & =y_{j}^{+}-2\sum_{k=1}^{N}\Theta(x_{k}-y_{j}^{+}+2N-k+2j-1),
\end{align}
($\Theta$ is the unit-step function) and
\begin{equation}
|\mathbf{x}_{N},\mathbf{y}_{M}^{-},\mathbf{y}_{M}^{+}\rangle=\left[\prod_{i=1}^{N}\sigma_{x_{i}}^{x}\right]\left[\prod_{i=1}^{M}\prod_{j=0}^{y_{i}^{+}-y_{i}^{-}}\sigma_{y_{i}^{-}+j}^{-}\right]|0\rangle.
\end{equation}

We can now easily calculate the Bethe equations for the particles. Let us pick up a particle and move it through the chain.
It will hit every other particle and bound state two times. The single time scattering
phases are $-e^{-i(p_{j}\mp p_{k})}$ and $-e^{-2ip_{j}}$ for magnons and
bound states. We can see that the momenta $p_k$ drop out from the Bethe equations since the products of the  scattering
phases
are $e^{-i(p_{j}-p_{k})}e^{-i(p_{j}+p_{k})}=e^{-i2p_{j}}$, and therefore the Bethe equations are
\begin{equation}
 e^{2ip_{j}(L-N-2M+2)}=1,
\end{equation}
where $N$ and $M$ are the number of the magnons and the bound states. We
can see that the Bethe equations are decoupled and they are the same as the one
magnon Bethe equation with length $\tilde{L}=L+1-N-2M$. Therefore
we obtained an effectively free theory with a modified volume.

We can check that the above defined states span the entire Hilbert space $\mathcal{H}_{\uparrow\uparrow}$. Fixing the number
of particles and bound states it is obvious that we can place the bound states in several
ways. The number of these possibilities is
\begin{equation}
\mathrm{DW}(L,N,M)= \binom{L-2N+2-2M}{2M}.\label{eq:deg}
\end{equation}
The effective length for the magnons is $\tilde{L}=L+1-N-2M$, therefore
the number of the solutions of the Bethe equations is
\begin{equation}
\mathrm{BA}(L,N,M)=\binom{L+1-N-2M}{N}. \label{eq:degBA}
\end{equation}
From \eqref{eq:deg} and \eqref{eq:degBA} we can count number of states we described above
\begin{equation}
 \Sigma(L) =
\sum_{N=1}^{(L+1)/2}\sum_{M=1}^{(L-2N+2)/4}\mathrm{DW}(L,N,M)\mathrm{BA}(L,N,M).
\end{equation}
We can convince ourselves that
\begin{equation}
\Sigma(L)=2^{L}=\dim\mathcal{H}_{\uparrow\uparrow}.
\end{equation}
We can see that we created as many states as the dimension of the Hilbert
space. We checked numerically that the free Bethe equations above with
the bound state degeneracies as given above produce the complete spectrum of the open
spin chain up to $L=8$.

\section{Bond-site transformation}

\label{sec:bond}

The goal of this Section is to build a different representation of the same model, such that the standalone Domain Walls
can be interpreted as particles (of a new particle species). We perform a non-local transformation: we put dynamical variables on the bonds between
sites, and build a dynamical model for the bonds. It turns out that our original charges $Q_4$ and $Q_2$ can be
represented by local operators after the transformation.
The advantage of such a representation is that it leads to a simpler Bethe Ansatz
description, with only two particle types. This Bethe Ansatz naturally respects the spin reflection invariance of the
original model. However, the advantages come at a cost: the original charge $Q_1$ becomes a non-local operator in the
new basis.

We put variables onto the bonds (links) between the sites, and we perform a change of basis from the old computational
basis to the new one. We have two options for each link, which we denote by $\circ$
and $\bullet$. We put $\circ$ if the two neighbours are of the same spin, and we put $\bullet$ if
the two neighbours are different. This transformation is completely invariant with respect to a global spin flip, which is an
invariance of the operators $Q_2$ and $Q_4$. This transformation is identical to the one used in
\cite{folded1,folded2} to derive their dual Hamiltonian; this is most easily seen from footnote 4 on page 11. of
\cite{folded1}, where the transformation rule for the $\sigma^z$ operators is given. The interpretation as a bond model
is new.

We can define this bond model on a periodic lattice, but for simplicity we first consider the boundary setting. We
consider the sector of the original model where the spin at site $j=1$ is in the up spin position. Then we construct the
bond basis for the $L'=L-1$ bonds, without any restriction on the last site. This way the Hilbert space will have a
dimension of $2^{L'}$.

Let us now give the operator representation of the charges in the bond basis.
In terms of spins, we can interpret $\ket{\circ}$ as the up spin, and
$\ket{\bullet}$ (the particle) as the down spin, and below we will use the Pauli matrices in this new basis accordingly.

Under this transformation the charges $Q_{2,3,4}$ become the local operators
\begin{equation}
  \label{bond-charges}
  \begin{split}
    Q^B_2&=\sum_{j=1}^{L'} \frac{1-\sigma^z_j}{2},\\
    Q^B_3&=\frac{i}{2}\sum_{j=1}^{L'-1}  \left(\sigma^+_j P^\bullet_{j+1}\sigma^-_{j+2}-  \sigma^-_j P^\bullet_{j+1}\sigma^+_{j+2} \right),  \\
    Q^B_4&=\frac{1}{2}\sum_{j=1}^{L'-1}  \sigma^-_j P^\bullet_{j+1}\sigma^+_{j+2}+  \sigma^+_j P^\bullet_{j+1}\sigma^-_{j+2},
  \end{split}
\end{equation}
where $P^\bullet_j$ is the projector onto the state $\ket{\bullet}$ on site $j$.
This form of the charges is obtained by direct computation. They were already given in \cite{folded1,folded2} using a
transformation on the level of the operators.

Interestingly, we can build the non-Hermitian combinations $Q^B_4\pm iQ^B_3$, that describe propagation terms towards
the left or to the right only.

A disadvantage of the bond picture is that $Q_1$ given in \eqref{charges12} is not a local operator anymore. Instead,
it is given by the highly non-local expression
\begin{equation}
  Q^B_1=\frac{1}{2} \sum_{j=1}^{L'}\left[1-   \prod_{k=1}^j \sigma^z_j\right].
\end{equation}

As noted earlier, we can also define the bond model with the operators above, assuming periodic boundary conditions. In
this case the bond model is not completely equivalent to the original spin chain: for example, it allows the presence of
an odd number of domain walls, which were forbidden by definition in the old chain. Nevertheless, the sectors with an
even number of DW's are equivalent to those of the original chain. In these sectors every state of the bond model
actually describes two different states from the original chain, which are related by spin reflection.

Sectors of the bond model with an odd number of DW's can be accommodated in the original spin chain if twisted boundary
conditions are chosen, with a twist given by spin reflection. Such a model is also interesting on its own right \cite{batchelor-baxter-antiperiodic,yung-batchelore-antiperiodic}, but we do not
discuss it here.

\subsection{Dynamics and Bethe Ansatz in the bond picture}

In the bond picture a single $\bullet$ represents an original DW, while two bullets on neighbouring sites represent an
original particle. Correspondingly, the kinematical terms in $Q^B_3$ and $Q^B_4$ move the double bullets, but they leave
the single $\bullet$ invariant. To be more precise, the only non-vanishing matrix elements of $Q^B_3$ and $Q^B_4$ are
those corresponding to the moves
\begin{equation}
  \label{bond-rules}
  \ket{\circ\bullet\bullet}\to \ket{\bullet\bullet\circ},\qquad
 \ket{\bullet\bullet\circ} \to
\ket{\circ\bullet\bullet}.
\end{equation}
We can regard the DW as the fundamental excitation, and the original particle (the doublet
$\bullet\bullet$) as a bound state of two DW's. The phenomenon that a single excitation is stable but a bound state of
two excitations is dynamical is similar to the situation in fracton models (see the reviews
\cite{fracton-review1,fracton-review2}).

Let us now construct the Bethe Ansatz wave functions in the bond
model. We set $h=0$ (thus we discard the non-local charge $Q_1^B$) and consider the local Hamiltonian
\begin{equation}
  H^B=Q^B_4+\mu Q^B_2.
\end{equation}
with periodic boundary conditions. The construction below is not rigorously proven: we construct the Bethe Ansatz by assuming factorized scattering and using the
$S$-matrix factors extracted from the two body problem. We put forward that the Bethe Ansatz is not unique: different
constructions lead to different choices for the basis vectors in the highly degenerate subspaces. We discuss this issue
at the end of this Section. 

We have two types of excitations in the model:
the single $\bullet$ which is a DW, and the $\bullet\bullet$ which is a particle. We will use the notations $p$ and
$DW$.

Correspondingly we introduce local creation operators
\begin{equation}
  \label{Aa}
  A^a_j=
  \begin{cases}
      \sigma^-_j & \text{ if } a=DW\\
    \sigma^-_j\sigma^-_{j+1} & \text{ if } a=p.\\
  \end{cases}
\end{equation}
Note that we have automatic exclusions:
\begin{equation}
  A^p_xA^p_{x+1}=A^p_xA^{DW}_{x+1}=0.
\end{equation}

Let us consider a state with $N$ particles and $M$ DW's; the set of the particle and DW momenta will be denoted as $\pn$
and $\km$. The wave function is most easily written down by merging these sets. Therefore we introduce a set of momenta
${\bf q}_{N'}$ and a set of particle types ${\bf a}_{N'}$ with $N'=N+M$.
We assume that there are no coinciding rapidities. 

The wave function is then written as
\begin{equation}
  \label{bondpsi}
  \begin{split}
      \ket{\Psi}=
  \sum_{x_1\le x_2\le \dots \le x_{N'}}
 \sum_{\mathcal{P}\in S_{N'}} e^{i \sum_{j=1}^{N'} q_{\mathcal{P}_j} x_j}  \\
\times \mathop{\prod_{j<k}}_{\mathcal{P}_j>\mathcal{P}_k}
S_{a_j,a_k}(q_{j},q_{k})
\prod_{j=1}^{N'} A^{a_{\mathcal{P}_j}}_{x_j} \ket{\emptyset}.
 \end{split}
\end{equation}
The scattering factors for different particle pairs are:
\begin{equation}
  \begin{split}
    S_{DW,DW}(q_1,q_2)&=S_{p,p}(q_1,q_2)=-e^{-i(q_1-q_2)},\\
    S_{p,DW}(q_1,q_2)&=e^{-i(q_1-2q_2)}.
 \end{split}
\end{equation}
For simplicity we assume that the original ordering of particle types is $2,2,2,\dots,2,1,1,1,\dots,1$.

Note that the $S$-matrix is such that for DW's
the occupation of neighbouring sites is forbidden. This ensures that
we do not mistake two domain walls close-by with a particle.
Also, if we have two particles at positions $x_1<x_2$, then $x_2=x_1+1$ is forbidden by the action of
the creation operators, but the next possibility $x_2=x_1+2$ is allowed.
Furthermore, if we have a DW at $x_1$ and a particle at $x_2$ then $x_2=x_1+1$ is allowed, but the other ordering
$x_1=x_2+1$ is forbidden. Thus, in this wave function a sequence of $\ket{\bullet\bullet\bullet}$ embedded in the vacuum
is interpreted as a DW and a particle from the left to the right. It
is merely a choice which follows from our definition of the creation
operators.

The energy of this state is
\begin{equation}
  E=\mu N_{DW}+  \sum_{a_j=p} (e(p_j)+2\mu).
\end{equation}
The sum runs over the particles only. The domain walls only contribute
to the energy through the chemical potential $\mu$.

The Bethe equations for the $p$ and $k$ variables are
\begin{equation}
  \label{bondBetheeq}
  \begin{split}
    e^{ip_jL}\mathop{\prod_{l=1}^{N}}_{l\ne j} S_{p,p}(p_j,p_l)
    \prod_{m=1}^M S_{p,DW}(p_j,k_m)&=1,\\
    e^{ik_lL} \mathop{\prod_{m=1}^M}_{m\ne l} S_{DW,DW}(k_l,k_m)
    \prod_{j=1}^N S_{DW,p}(k_l,p_j)&=1.
  \end{split}
\end{equation}
Substituting the factors we get
\begin{equation}
  \label{bondBetheeq2}
   \begin{split}
    e^{ip_j(L-N-M)} e^{iP} e^{2iK} &=(-1)^{N-1},\\
    e^{ik_l(L-2N-M)} e^{iK}e^{iP}&=(-1)^{M-1},
  \end{split}
\end{equation}
where we defined
\begin{equation}
  P=\sum_{j=1}^Np_j,\qquad K=\sum_{j=1}^Mk_j.
\end{equation}
The energy is carried only by the particles, and the effect of the domain walls
is only a change in the available volume. Correspondingly, the actual values of the $k_j$ variables do not matter for
the particle momenta $p_j$, which is influenced only by the sum $K$. Thus the distribution of $K$ among the variables
$k_j$ only contributes to the degeneracies of the energy levels. {This can be seen more explicitly by taking the product of
the second set of equations, which leads to the following coupled equations for the variables $p_j$ and $K$:
\begin{equation}
  \label{bondBetheeq3}
   \begin{split}
    e^{ip_j(L-N-M)} e^{iP} e^{2iK} &=(-1)^{N-1},\\
    e^{iK(L-2N)}e^{iPM}&=1,
  \end{split}
\end{equation}}
Explicit solutions to the equations above are found as follows. The overall momenta are expressed as
\begin{equation}
\begin{split}
&P\frac{L(L-M-2N)}{(L-2N)}\\
&=\pi\left(N-1-\frac{2M(M-1)}{L-2N}\right)N+2\pi I-4\pi \frac{JN}{L-2N},
\end{split}
\end{equation}
and
\begin{equation}
K=\frac{\pi(M-1)M+2\pi J-PM}{L-2N},
\end{equation}
where $I$ and $J$ are arbitrary integers.
Afterwards the particle momenta $p_j$ are expressed as
%
%\begin{multline}
%p_j=\frac{\pi}{L-N-M}\left[\left(N-1-\frac{2M(M-1)}{L-2N}\right)+2I_j\right]
%\\-\frac{\pi}{L-N-M}\frac{P(L-2N-2M)+4JN}{(L-2N)},
%\end{multline}
%
%\begin{multline}
%p_j=\frac{\pi}{L-N-M}\left[N-1+2I_j-\frac{2M(M-1)+4J}{L-2N}\right]
%\\-\frac{P}{L-N-M}\frac{L-2N-2M}{L-2N},
%\end{multline}
%or simply just
\begin{equation}
p_j=\frac{(N-1+2I_j)\pi-P-2K}{L-N-M},
\end{equation}
where  quantum numbers $I_j$ are given by $I=\sum_{s=1}^N I_s$:
$0<I_1<I_2\dots<I_N<L$.
%and the condition $i(P+K)L=0 \mod(2\pi)$
%has to be satisfied.
%

Let us compare these equations with
\eqref{S1n} that was derived using the strings of the XXZ model.
We can see in \eqref{S1n}
that the contribution to the volume change is the same
for every string, thus for every bound state. In the bond picture this means that the volume change does not
depend on the relative position of the DW's, just their total
number. This is consistent with the equations above. However, the
``twist'' felt by the particles depends on the momentum of the DW's.

This basis is certainly different from the one obtained by the string
picture in the original model. The key difference is that here the
DW's are allowed to move independently, whereas in the string picture
the two DW's of the bound state are always at a fixed
distance. Both pictures describe highly degenerate energy
levels, and such differences only amount to a free choice of the
basis.

In fact, it is easy to see that the Bethe states obtained in this bond picture can not be identical to those of the
original model. This happens because in the bond picture the two DW's associated to an original string solution can not
have the same momenta, therefore they can not move together. Furthermore, the wave function \eqref{bondpsi} naturally
gives a linear combination of states where the ``string lengths'' vary. States with fixed string lengths are not
reproduced by this formula. However, we stress that the two descriptions merely correspond to two different choices for the basis vectors in a highly
degenerate eigenspace.

It is worth mentioning that the Bethe Ansatz solution presented above is completely different from the one of
\cite{folded1,folded2}, and it is not straightforward to find a dictionary between the two solutions, and most probably
they  lead to different basis vectors in the highly degenerate eigenspaces. We checked by numerical computations
that both constructions correctly reproduce the spectrum up to $L=8$.

\section{Non-local mapping to the $SU(3)$ XX model}

\label{sec:MM}

Here we show that the bond model can be mapped to the
Maassarani-Mathieu (MM)
spin chain, which is also known as the $SU(N)$ XX model \cite{su3-xx}; our mapping concerns the $SU(3)$-related case.
The mapping is similar in spirit to the non-local mapping between the phase model and the XX model which
first appeared in \cite{qboson-bog3} and which was treated detail in \cite{sajat-q2}.
The mapping for the present model is new, and it is different from the connection with the Bariev model \cite{bariev-model} found in
\cite{folded1}.

We start with the bond model with length $L$ in the boundary setting. This model will be mapped to a spin chain with
local dimension 3, with local basis states denoted as $\ket{1}$, $\ket{2}$ and $\ket{3}$. As before, we construct the
mapping on the level of the basis states. We will see that the mapping is volume changing: different states are mapped
to states of the new model with different lengths.

The rules for the mapping in the computational basis are as follows. We represent each basis state as a sequence of
$\bullet$'s and $\circ$'s. We proceed from the left to the right of this sequence and we translate it to a new
sequence consisting of the numbers $1,2,3$. If at a given position we encounter a $\circ$ then we add a $1$ to the new
sequence. Then we move further to the next entry. If we encounter a $\bullet$ then we also need to check the next entry:
a pair $\bullet\bullet$ is then mapped to $2$, whereas the pair $\bullet\circ$ is mapped to $3$. This rule is then
applied as we proceed along the chain.

This mapping works flawlessly on a half-infinite chain, but it runs into a problem
on a finite chain if the last entry is a single $\bullet$ for which no rule is specified. Here we are not concerned with
the boundary conditions, we are focusing on the mapping on the bulk, thus we discard this problem. It is possible that
an exact mapping with open boundary conditions could be constructed, but we leave this problem to a future work.

The mapping backwards is more easily summarized as
\begin{equation}
   \ket{1}\to\ket{\circ},\qquad\ket{2}\to\ket{\bullet\bullet},\qquad \ket{3}\to\ket{\bullet\circ}.
\end{equation}

Let us discuss the changes in the length. If the numbers of the components $1,2$ and $3$ in the new sequence are given
by $N_{1,2,3}$,
with the total  length being
\begin{equation}
 L'=N_1+N_2+N_3,
\end{equation}
then the length of the original chain is
\begin{equation}
L=N_1+2N_2+2N_3.
\end{equation}
We see that different basis states of our model are mapped to MM chains with different length. It is useful to note that
the original particle and DW numbers are
\begin{equation}
 N=N_2,\qquad M=N_3,
\end{equation}
thus the volume of the MM chain for a given $N,M$ reads
\begin{equation}
  \label{Lprime}
  L'=L-N-M.
\end{equation}

We focus on the Hamiltonian in the bond model given by $Q^B_4$ in \eqref{bond-charges}, which results in the transition
rules \eqref{bond-rules}. It can be checked that after the basis transformation this translates into the
transition matrix elements
\begin{equation}
  \ket{12}\leftrightarrow \ket{12},\qquad \ket{23}\leftrightarrow \ket{23}.
  \label{MM:transitionmatrix}
\end{equation}
We also write down the Hamiltonian that encodes  these transitions.  Let $E^{a,b}_j$ with $a,b=1\dots 3$ be the
elementary matrices that contain a single 1 in row $a$ and column $b$, acting on the local Hilbert space on site $j$ of
the new model. Then the MM Hamiltonian is \cite{su3-xx}
\begin{equation}
 H_{\rm MM}=-\frac{1}{2} \sum_{j=1}^{L'} \sum_{a=1}^2 \left[E^{a,a+1}_jE^{a+1,a}_{j+1}+E^{a+1,a}_jE^{a,a+1}_{j+1}\right].
 \label{eq:MM}
\end{equation}
This is now written down for periodic boundary conditions and we included a factor of -1/2 to match our previous
normalizations. We note that this model is also equivalent to the infinite interaction limit of the Hubbard model (also
known as the t-0 model), see
for example \cite{maassarani-mathieu-hubbard,Hubbard-Book}.

It is useful to consider the specific sector of the bond model, which includes only particles. This sector is equivalent
to the
constrained XXZ model treated in Sec. \ref{sec:constrained}. Applying the non-local mapping to this sector we see that
only the local states $1$ and $2$ are populated. In this sector the MM chain is equivalent to the standard XX model.  In
\cite{sajat-q2} it was shown that the so-called phase model is equivalent to the XX model by a similar non-local
transformation. We have thus obtained three different models that are equivalent to each other via the non-local
mapping.

We stress that the equivalence of these models is established only for the bulk, because generally the boundary
conditions spoil the mapping. This situation is analogous to the one treated in \cite{sajat-q2}. As mentioned above, an
exact mapping could be found perhaps with open boundary conditions, but we do not pursue this direction here.

We now briefly discuss the spectrum of the MM Hamiltonian \eqref{eq:MM} with periodic boundary conditions, and the
construction of its eigenstates through Bethe Ansatz. An algebraic treatment was presented in
\cite{su3-xx}, and here we discuss the coordinate Bethe Ansatz solution. In fact we will see that there are two
different constructions, which are related to each other by a particle-hole transformation.

The first version of the coordinate Bethe Ansatz is close in spirit to the usual nested Bethe Ansatz, where we start
with a chosen reference state and build excitations above it, such that the orientation of the excitations will involve
the ``nesting'', or second Bethe Ansatz.
%The Hamiltonian \eqref{eq:MM} commutes with the total numbers $N_1,N_2,N_3$ of local states of each type.
We start with the sector
where only the local states $\ket{1}$ and $\ket{2}$ are populated; in this sector the MM model is equivalent to the
standard XX model. The eigenstates in this sector are therefore constructed as excitations on top of the pseudo-vacuum
$\ket{\rm vac} = \ket{2\ldots 2}$. They are parameterized by a set of momenta $\mathbf{k}=\{k_1,\ldots k_N\}$, and take
the form
\begin{equation}
\ket{\mathbf{k}  } =
\sum_{x_1 <  \ldots < x_N}
\sum_{\mathcal{P} \in \mathfrak{S}_N}
\mathrm{sgn}(\mathcal{P}) e^{i \sum_{j=1}^N k_{\mathcal{P}_j} x_j }
\ket{1_{x_1}\ldots 1_{x_N}}  \,,
\label{eq:CBAMM1}
\end{equation}
where we have used the notation $\ket{1_{x_1}\ldots 1_{x_N}} = E_{j_1}^{2,1}\ldots E_{j_n}^{2,1}\ket{\rm vac}$.
Periodicity of the wave function then imposes the quantization condition $e^{i k_j L'} = (-1)^{N-1}$ for each $j$, and for a given value of $N$ the momenta $k_1,\ldots k_N$ may take any of the $L$ distinct values compatible with this condition, provided that they are all distinct.
The corresponding energies are given by
\begin{equation}
E = \sum_{j=1}^N \cos(k_j).
\label{eq:CBAMMenergies}
\end{equation}
Turning to more general sectors, that is, with $N_3\neq 0$, we introduce the states $\ket{\mathbf{k}^{\alpha_1, \ldots, \alpha_{N}}}$, with $\alpha_j \in \{1,3\}$. These have the same form as \eqref{eq:CBAMM1}, but where the sequence $1_{x_1}\ldots 1_{x_N}$ is replaced by the corresponding sequence of 1s and 3s (we then have $N=N_1+N_3$).
Since the order of 1s and 3s along the chain is not modified by the transition matrix elements
\eqref{MM:transitionmatrix}, the states $\ket{\mathbf{k}^{\alpha_1, \ldots, \alpha_{N}}}$ are locally eigenstates of the
MM Hamiltonian, with energies again given by \eqref{eq:CBAMMenergies}. However, on a chain of finite size requiring
periodicity of the wave function mixes states with sequences ${\alpha_1, \ldots, \alpha_{N}}$ related through cyclic permutations. We therefore construct eigenstates as linear combinations of the form
\begin{equation}
\sum_{n=1}^N e^{i q n} \ket{\mathbf{k}^{U^n(\alpha_1, \ldots, \alpha_{N})}} \,,
\label{eqref:CBAMMaux}
\end{equation}
where $U$ cyclically permutes the indices $\alpha_i$. The pseudo-momentum $q$ is quantized through $e^{i q N} = 1$, but
depending on the sequence $\alpha_1, \ldots, \alpha_{N}$ only a subset of the solutions for $q$ might give rise to a
non-vanishing wave function. In other terms, constructing the linear combinations \eqref{eqref:CBAMMaux} amounts to
diagonalizing the one-site translation operator on an auxiliary spin-1/2 chain of $N$ sites, in the sector with $N_1$
spins $\uparrow$ and $N_3$ spins $\downarrow$.

Imposing the periodicity of the wave function, we now have the following quantization conditions
         \begin{equation}
        \begin{split}
               e^{i k_j L'} &= (-1)^{N-1}  e^{i q}, \,~~~  j=1, \ldots N,\\
      e^{i q N} &= 1.
     \label{eqref:CBAMMfinal}
      \end{split}
      \end{equation}
In summary, the eigenstates in a sector of given $N_1, N_2, N_3$ are obtained by first, finding the allowed values of $q$ by diagonalizing an auxiliary 
spin-1/2 problem, and, second, solving the quantization condition \eqref{eqref:CBAMMfinal} for the momenta $k_1, \ldots k_N$.
This is a particularly simple form of nested Bethe Ansatz, as the quantization condition of the auxiliary momenta $q$
does not depend on the values of the momenta $k_1, \ldots k_N$. We checked against exact diagonalization for finite size
systems that this construction indeed reproduces the entire spectrum of the periodic MM model.

Now we turn to the second construction for  the coordinate Bethe Ansatz, which is in fact closer in
spirit to the bond model presented in the previous section.
Rather than using a unique pseudo-vacuum, now we start with a collection of pseudo-vacua $\ket{(\alpha_1, \ldots \alpha_{N})}$ made of arbitrary sequences of 1s and 3s, which are all zero energy eigenstates of the MM Hamiltonian. On top of a pseudo-vacuum $\ket{(\alpha_1, \ldots \alpha_{N})}$, we create excitations by inserting local states 2 in between the 1s and 3s. The resulting Bethe wave functions take the form of the usual XX wave functions on a chain of $L'=N_2+N$ sites, analogous to \eqref{eq:CBAMM1} but where the locally excited sites $1_{x_i}$ are now replaced by $2_{x_i}$, and the states on other sites are distributed according to the sequence $(\alpha_1, \ldots \alpha_{N})$.
As in the previous construction such wave functions are locally eigenstates of the Hamiltonian, but periodicity now mixes
excitations on top of different pseudo-vacua, namely related by cyclic permutations. Eigenstates of \eqref{eq:MM} are
therefore obtained as linear combinations of excitations over cyclically permuted pseudo-vacua, involving as before a
pseudo-momentum $q$ with is a $N$th root of unity. The resulting quantization conditions for the momenta $k_j$ of the
``2'' particles take an analog form to \eqref{eqref:CBAMMfinal}, more precisely the two sets are mapped onto each other
by a particle-hole transformation of the underlying XX model.  Note the key difference that now the local states 2 are
the excitations, whereas they formed the reference state in the previous construction.

It is useful to compare \eqref{eqref:CBAMMfinal} to the Bethe equations obtained in the bond model, see
eq. \eqref{bondBetheeq3}. We observe some similarities, for example the apparent volume for the $p$-variables is the
same. However, the twists appearing in those equations are different. This is a consequence of the
fact that the non-local mapping is not compatible with the periodic boundary conditions in these models.

\section{Degeneracies and Hilbert space fragmentation}

\label{sec:deg}

Here we discuss two closely related features of the models: Hilbert space fragmentation and a large number of
degeneracies present in the spectrum. These questions were already discussed in \cite{fracton1,folded1,folded2}, here we
summarize the key statements and present a complementary view of the matter.

The expression ``Hilbert space fragmentation'' means that there are a large number of sectors in the Hilbert space such
that the Hamiltonian does not have transition matrix elements between the sectors. It is also required that the sectors
should be constructed using relatively simple rules, before actually solving the full dynamics in the model. In a
typical case there is an exponentially large number of disconnected sectors in the Hilbert space. Fragmentation is known
to happen in models with fractonic excitations  \cite{fracton-review1,fracton-review2}.

One of the mechanisms for the fragmentation is the presence of conserved charges \cite{tibor-fragment,fragment-fracton-2}.
If there is
a $U(1)$ conserved charge of the model, then its eigenvalues already split the Hilbert space into various sectors, however, such a
splitting is very common and natural, and it is usually not called ``fragmentation''. In contrast, fragmentation can happen
in the presence of two conserved charges with local densities, such that the charges are not dynamical and their
densities can be diagonalized simultaneously. Examples  with a charge and a dipole conservation were discussed in
\cite{tibor-fragment,fragment-fracton-2}.

In our case we have two such charges $Q_1$ and $Q_2$, which commute with the dynamical charge $Q_4$. Accordingly, the
eigenvectors organize themselves  into sectors corresponding to the eigenvalues $\Lambda_{1,2}$ of the charges $Q_{1,2}$
Furthermore, we observe that even
the sectors with a given eigenvalue pair $(\Lambda_1,\Lambda_2)$ further split into sub-sectors, whose number grows exponentially with the
volume.  These sectors correspond to the presence of the domain walls placed at various distances from each other.

An intuitive way to understand the phenomenon is to first consider the completely frozen states which have an eigenvalue 0
under $Q_4$. As explained above, such states can be created by placing an arbitrary number of domain walls on top of a
reference state with minimum distances of at least 2. Then the particle excitations are created above such a frozen
state, such that the Hilbert space remains fragmented into these sectors.

Despite the simplicity of this picture, it is not completely precise in a finite volume situation with periodic
boundaries: Even though the
domain walls themselves are frozen, they are displaced once a scattering with a particle occurs. Therefore, we need to
impose the proper periodicity conditions also for the movement of the DW's, resulting in the full set of Bethe
equations. In contrast, the wave function given in Section \ref{sec:boundary} shows that in the boundary case the fragmentation
can be understood using this intuitive picture, because it is relatively easy to treat the displacements of the bound
states.

Let us now also discuss the pattern of degeneracies in the model. We study the most general Bethe Ansatz equations
\eqref{generalBA}. First we discuss the degeneracies at $h=0$. The energy is carried only by the particles, thus it
depends only on the set ${\bf p}_{N_1}$, and the global quantum numbers $N_1$ and $N_s$. The Bethe equations for
${\bf p}_{N_1}$ do not depend on the distribution of string lengths; they are sensitive only to the total string number
$N_s$ and the total string momentum. Thus we observe a large degeneracy, which is expected to be exponentially growing
with the system size. In the bond picture this degeneracy results from an arbitrary placement of the DW's, which does
not affect the quantization conditions for the particles.

The exponential growth of the degeneracies strongly depends on the particle content of a
state. Each domain wall decreases the space available for the propagation of the particles, thus a given particle number
$N_1$ also constrains the possibilities for the DW's; this was explicitly demonstrated in the boundary case in Section
\ref{sec:boundary} above. For the reference state the
exponential growth was computed in \cite{fracton1} and also in \cite{folded1}, and it was found that the degeneracy behaves in large volumes as
$\alpha^L$, where $\alpha=(1+\sqrt{5})/2$ is the golden ratio. In contrast, the ground state has a finite degeneracy
even in the thermodynamic limit (see below).

Let us also discuss the splitting of the energy levels as a magnetic field is switched on. The additional term in
\eqref{Egen} implies that most of the degeneracies are split due to the various distributions of the total string number
$N_s$ into the different strings. However, an exponential amount of degeneracy still remains, corresponding to the
various ways of obtaining the same $N_s$ and the same total magnetization, 
and also to the various ways of solving the Bethe equations for the strings.

%We first discuss degeneracies
%at $h=\mu=0$, and then also explain how some of them are lifted when $h\ne 0$ and/or $\mu\ne 0$.

\section{Ground state}

\label{sec:gs}

Here we discuss the nature of the ground state of the original Hamiltonian given by \eqref{H} and we compute the ground
state energy density. We consider the cases
$h=\mu=0$ and also the situations when either or both are switched on.

For the case of $h=\mu=0$ our results agree with
those of \cite{folded1} and in the earlier work \cite{constrained1}. Simple arguments and numerical checks show that the ground state is populated by
particles only, but their overall density (or equivalently, the Fermi boundary) is a non-trivial quantity. The ground
state is doubly degenerate, corresponding to the spin flip invariance, and here we treat the state that has positive
overall magnetization ($N<L/2$).

The Bethe equations are:
\begin{equation}
e^{ip_jL}\prod_{k=1}^N\big(-e^{-i(p_j-p_k)}\big)=-1.
\end{equation}
Taking the logarithm, we get
\begin{equation}
p_jL=(N-1)\pi+\sum_{k=1}^N(p_j-p_k)+2\pi I_j,
\end{equation}
where $I_j\in \mathbb{Z}$. Since in the ground state $P=\sum_{k=1}^Np_k=0$, this simplifies to
\begin{equation}
p_j=\frac{2\pi}{L-N}\tilde{I}_j.
\end{equation}
Here $\tilde{I}_j=I_j+(N-1)/2$ is an integer/ half integer for odd/even $N$. In the ground state the $\tilde{I}_j$-s are distributed symmetrically around zero, from $-\frac{N-1}{2}$ to $\frac{N-1}{2}$. The energy of the state is
\begin{equation}
E=-\sum_{j=1}^N\cos\left(\frac{2\pi}{L-N}\tilde{I}_j\right).
\end{equation}
In the thermodynamic limit ($L,N\rightarrow\infty,\ N/L=n$ is fixed) the energy density becomes:
\begin{multline}
  \epsilon=\frac{E}{L}=-\int_{-p_F}^{p_F}dp\frac{L(1-n)}{2\pi L}\cos(p)\\
=-\frac{1-n}{\pi}\sin(p_F)=-\frac{1-n}{\pi}\sin\left(\frac{n\pi }{1-n}\right),
\end{multline}
where we used that
\begin{equation}
  p_F=\frac{2\pi}{L-N}\cdot\frac{N-1}{2}\quad\rightarrow\quad  \frac{n}{1-n}\pi.
\end{equation}
The particle density that minimizes $\epsilon$ is found from
\begin{equation}
\frac{\partial\epsilon}{\partial n}=0,
\end{equation}
which leads to
\begin{equation}
\frac{1-n}{\pi}\tan\left(\frac{n\pi}{1-n}\right)=1.
\end{equation}
This equation can be solved numerically and gives $n\approx 0.3008$. The energy density is  $\epsilon\approx -0.2172$.

If we include bound states, their only effect for the particles is decreasing the effective length by $2M$, where $M$ is the number of bound states (with arbitrary length). Therefore the energy density in a state with particle density $n$ and bound state density $M/L=m$ is:
\begin{equation}
\epsilon=-\frac{1-n-2m}{\pi}\sin\left(\frac{n\pi }{1-n-2m}\right).
\end{equation}
This expression has its minimum at $n=n_0\approx 0.3008$ and $m=0$. Since the $Q_2$ term gives the same energy to a
particle and a bound state and the $Q_1$ term energetically prefers particles over bound states, the situation will not
change even if we turn on non-zero $h$ and $\mu$. The ground state does not contain bound states.

Let us now investigate the case with $h\neq 0$ but still $\mu=0$. We can restrict ourselves to $h>0$ by spin reflection
invariance.
In the thermodynamic limit the energy density of the system in a state characterized by a particle density $n$ is
\begin{equation}
\epsilon=-\frac{1-n}{\pi}\sin\left(\frac{n\pi }{1-n}\right)+hn.
\end{equation}
The value of $n$ that minimizes this energy density is given by
the relation
\begin{equation}
h=\frac{1}{1-n}\cos\left(\frac{n\pi }{1-n}\right)-\frac{1}{\pi}\sin\left(\frac{n\pi }{1-n}\right) \label{gs_h_n}.
\end{equation}
The expression on the r.h.s. has a maximum at $n=0$ with a value of 1. This means that if $h<1$, the ground state is
characterized by a finite particle density $n_0$, given by \eqref{gs_h_n}. On the other hand, if $h>1$ than the
reference state with all spins up becomes the ground state. By expanding \eqref{gs_h_n} around 0, we can calculate how
$n$ goes to 0, as $h$ approaches $h_c=1$. We find the scaling
\begin{equation}
n\propto (h_c-h)^{1/2}.
\end{equation}

Now we consider the case of $h\neq 0, \mu\neq 0$. It is still enough to consider $h>0$, but $\mu$ can be both positive
and negative. The energy density in a state characterized by particle density $n$ is
\begin{equation}
\epsilon=-\frac{1-n}{\pi}\sin\left(\frac{n\pi}{1-n}\right)+(h+2\mu)n.
\end{equation}
By taking the derivative and rearranging the equation we get
\begin{equation}
h+2\mu=\frac{1}{1-n}\cos\left(\frac{n\pi}{1-n}\right)-\frac{1}{\pi}\sin\left(\frac{n\pi}{1-n}\pi\right).
\end{equation}
The expression on the r.h.s has a maximum at $n=0$ with a value of 1, and has a minimum at $n=1/2$ with a value of
-2. Therefore, if $h+2\mu>1$, then the reference state with all spins up becomes the ground state. On the other hand, if
$h+2\mu<-2$, the ground state will be given by $n=1/2$ which corresponds to the doubly degenerate N\'eel and anti-N\'eel states.
Between these two regions the ground state is characterized by a finite $n$, given by the following constraint:
\begin{equation}
\frac{1}{\pi}\sin\left(\frac{n \pi}{1-n}\right)-\frac{1}{1-n}\cos\left(\frac{n\pi }{1-n}\right)+h+2\mu=0.
\end{equation}

\section{Thermodynamics}

\label{sec:thermo}

Here we investigate the Gibbs states of the model, and compute the free energy and particle densities as a function of
the temperature and chemical potential. We present three  different computations using the various
formulations of the model, all leading to the same free energy density.

\subsection{Thermodynamics in the boundary case}

Here compute the thermodynamic limit of the boundary chain discussed in \ref{sec:boundary}.
In a certain sense this is the simplest case which leads to an exact determination of the free energy density, without
recourse to additional assumptions.

\begin{figure}
\begin{centering}
\includegraphics[width=0.95\columnwidth]{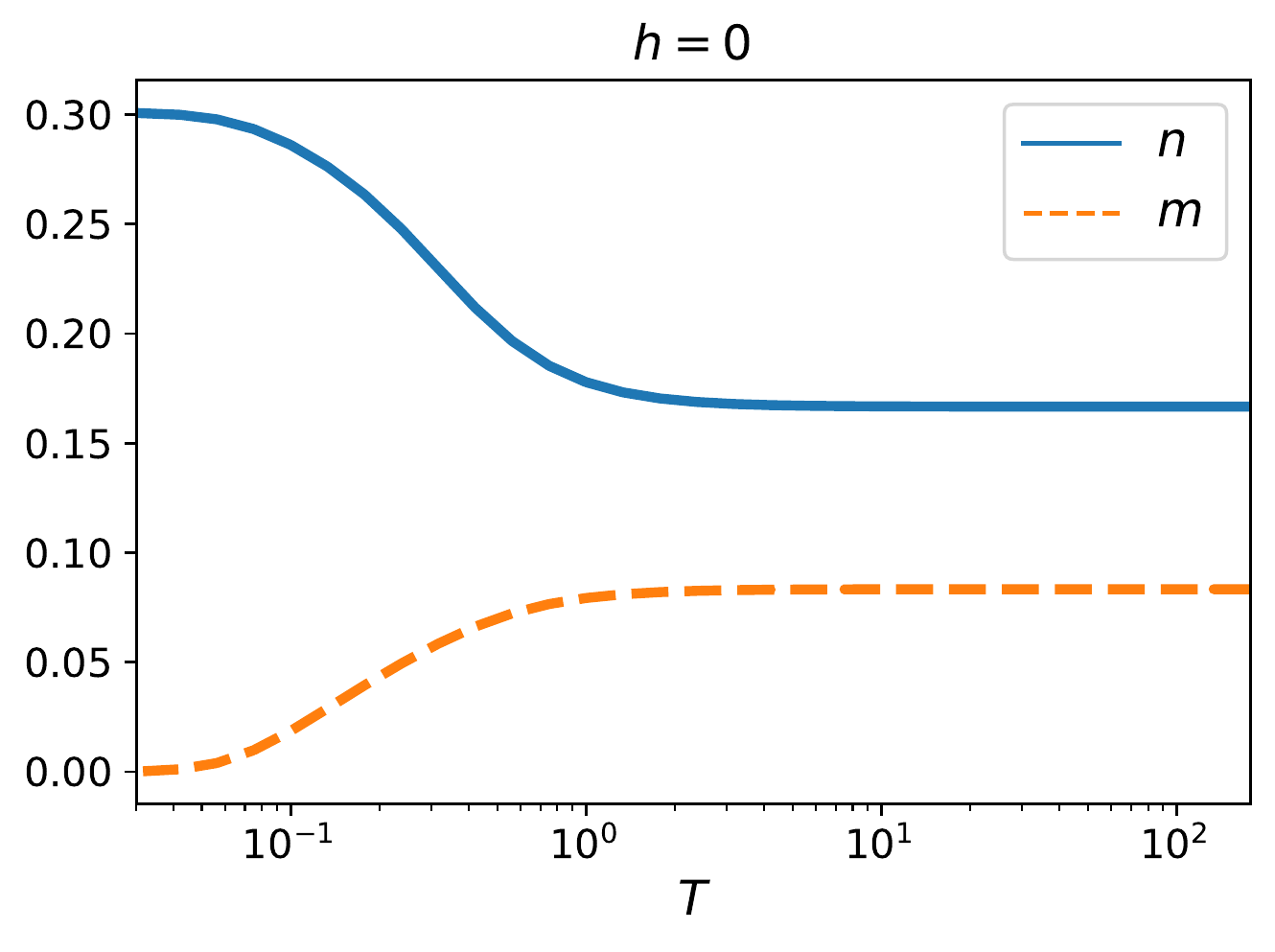}
\par\end{centering}
\caption{Thermodynamic limit of the densities $n=N/L$, $m=M/L$. The horizontal axis is the temperature $T$.}
\label{fig:TBAopen}
\end{figure}

For simplicity we focus on the $\mathcal{H}_{\uparrow\uparrow}$
sector of the open chain. We saw that the spectrum can be built from particles and
bound states. Let us define the particle density $\tilde\rho(p)$ for which $(L-N-2M)\tilde\rho(p)\Delta p$
gives the number of particles in the interval $\Delta p$. Note that his is an unconventional definition because it uses
the modified (apparent) volume.
Using this density the number of particles can be written as
\begin{equation}
N=(L-N-2M)\int_{0}^{\pi}\tilde\rho(p)dp.\label{eq:condN}
\end{equation}
In the thermodynamic limit ($L\to\infty$, $N/L=n$ and $M/L=m$)
the free energy can be written as
\begin{equation}
f[\tilde\rho,n,m]=E[\tilde\rho,n,m]-TS_{YY}[\tilde\rho,n,m]-TS_{DW}(n,m),
\end{equation}
where the energy term is
\begin{multline}
E[\tilde\rho,n,m]\\=\left(1-n-2m\right)\int_{0}^{\pi} e(p)\tilde\rho(p)dp
\end{multline}
with $e(p)$ given by \eqref{ep}.
The Yang-Yang entropy is
\begin{multline}
S_{YY}[\tilde\rho,n,m]=-\left(1-n-2m\right) \\
\times\left[\int_{0}^{\pi}\tilde\rho\log(\tilde\rho)+\left(\frac{1}{\pi}-\tilde\rho\right)\log\left(\frac{1}{\pi}-\tilde\rho\right)dp+\log\pi\right]
\end{multline}
and the entropy $S_{DW}$ comes from the degeneracy (\ref{eq:deg})
\begin{multline}
S_{DW}(n,m)=(1-2n-2m)\log(1-2n-2m)\\
-2m\log(2m)-(1-2n-4m)\log(1-2n-4m).
\end{multline}
Let us introduce a Lagrange multiplier for the particle number (\ref{eq:condN})
\begin{multline}
f[\tilde\rho,n,m,\alpha]=f[\tilde\rho,n,m]\\ -\alpha\left[(1-n-2m)\int_{0}^{\pi}\tilde\rho(p)dp-n\right].\label{eq:freeen}
\end{multline}
Taking the functional derivative with respect to $\tilde\rho$ we obtain that
\begin{equation}
\frac{\delta f}{\delta\tilde\rho}:\qquad\left(1-n-2m\right)\left(e(p)-T\log\frac{\frac{1}{\pi}-\tilde\rho}{\tilde\rho}-\alpha\right)=0,
\end{equation}
therefore we obtain the usual free fermion density where the Lagrange
multiplier acts as a chemical potential:
\begin{equation}
\tilde\rho(p)=\frac{1}{\pi}\frac{1}{e^{\beta(e(p)-\alpha)}+1}.
\end{equation}
The advantage of the multiplier is that we can change the functionals
to simple functions
\begin{equation}
S[\tilde\rho]\to S(\alpha).
\end{equation}
The particle numbers $n,m$ and a Lagrange multiplier are given by
\begin{align}
\frac{\partial f}{\partial\alpha}  :\quad0 =& (1-n-2m)\int_{0}^{\pi}\tilde\rho(p)dp-n,\\
\frac{\partial f}{\partial n}  :\quad0 = & f_0(\alpha,n,m)-(1-2m)\alpha \nonumber\\
&-2T(1-n-2m)\log\frac{1-2n-2m}{1-2n-4m},\\
\frac{\partial f}{\partial m}  :\quad0= & f_0(\alpha,n,m) -n\alpha \nonumber\\
-T&(1-n-2m)\log\frac{(1-2n-2m)2m}{(1-2n-4m)^{2}},
\end{align}
where
\begin{equation}
 f_0(\alpha,n,m)=E(\alpha,n,m)-TS_{YY}(\alpha,n,m).
\end{equation}
We can simplify the expression of $f_0$ as
\begin{multline}
 f_0(\alpha,n,m) =-T(1-n-2m) \\
\times\int_0^\pi \frac{dp}{\pi}\log \left( 1+e^{-\beta(e(p)-\alpha)} \right)+\alpha n.
\end{multline}
Therefore the solutions of the following equation system gives the
finite temperature ground state $(n,m,\alpha)$
\begin{align}
 \int_{0}^{\pi}\frac{dp}{\pi}\frac{1}{e^{\beta(e(p)-\alpha)}+1} &=\frac{n}{1-  n-2m},\label{eq:eqsys1}\\
  \alpha &=T\log\frac{2m}{1-2n-2m}\label{eq:eqsys2}
\end{align}
and
\begin{multline}
\int_0^\pi \frac{dp}{\pi}\log \left( 1+e^{-\beta(e(p)-\alpha)} \right) \\
= - \beta\alpha + 2 \log\frac{1-2n-4m}{1-2n-2m}.\label{eq:eqsys3}
\end{multline}
We solved this equation system numerically and the result is diaplayed on Figure
\ref{fig:TBAopen}. In the $T\to0$ limit it agrees with the ground
state result given in Section \ref{sec:gs}.
% i.e. $n\approx3.008,m\approx0$.

Substituting (\ref{eq:eqsys1}-\ref{eq:eqsys3}) into (\ref{eq:freeen})
we obtain that the free energy can be written as
\begin{equation}
 f/T=\log\frac{2m}{1-2n-4m}.
\end{equation}

If someone is interested only on the free energy and not on the values of $n$ and $m$ then the equation system simplifies to a single equation. Let us use the following variable
\begin{equation}
 x = e^{-\beta \alpha} = \frac{1-2n-2m}{2m},
\end{equation}
where we used \eqref{eq:eqsys2}. The equation \eqref{eq:eqsys3} can be rewritten as
\begin{equation}
 \int_0^\pi \frac{dp}{\pi}\log \left( x+e^{-\beta e(p)} \right) =
 2 \log(x-1).
\end{equation}
The free energy can be also expressed by $x$ as
\begin{equation}
 f/T=\log\frac{x}{x-1}.
\end{equation}

Let us continue with the $h\neq 0$ case.
Let $M_{k}$ be the number of DW with length $k$. Let us the following
notations
\begin{align}
M & =M_{2}+M_{3}+M_{4}+\dots\\
yL & =L-2N+2-4M_{2}-5M_{3}-6M_{4}-\dots
\end{align}
Furthermore we will use $\mathbf{m}=\{m_1,m_2,\dots\}$ with $m_j=M_j/L$.

The degeneracy of states with these quantum numbers is
\begin{equation}
\frac{(yL+M)!}{(yL)!M_{2}!M_{3}!M_{4}!\dots},
\end{equation}
therefore in the thermodynamic limit the DW entropy is
\begin{multline}
S_{DW}(n,\mathbf{m})=(y+m)\log(y+m)-y\log y \\
-m_{2}\log(m_{2})-m_{3}\log m_{3}-m_{4}\log m_{4}-\dots.
\end{multline}
The free energy is
\begin{multline}
f[\tilde\rho,n,\mathbf{m},\alpha]=E[\tilde\rho,n,m]-TS_{YY}[\tilde\rho,n,m]\\
-TS_{DW}(n,\mathbf{m})-
\alpha\left[(1-n-2m)\int_{0}^{\pi}\tilde\rho(p)dp-n\right]\\
+h(n+2m_{2}+3m_{3}+4m_{4}+\dots).
\end{multline}
We can see that the $\tilde\rho$ and the $\alpha$ dependent parts are the
same as before with the same constraint.
For the derivatives with respect to  $n,m_{k}$ we obtain the following equations
\begin{align}
\nonumber\frac{\partial f}{\partial n}:& \quad  f_0(\alpha,n,m)=
(1-2m)\alpha  \\
+ 2T&(1-n-2m)\log\frac{y+m}{y}+(1-n-2m)h,\\\nonumber
\frac{\partial f}{\partial m_{k}} :& \quad f_0(\alpha,n,m)= n\alpha  \\\nonumber
& +\frac{1}{2}T(1-n-2m)\log\frac{(m+y)^{k+1}m_{k}}{y^{k+2}}\\
& +(1-n-2m)\frac{k}{2}h.
\end{align}
Making proper subtractions we obtain
\begin{align}
 \alpha &= \frac{T}{2}\log\frac{m_{2}}{y+m}, \\
 m_{k+1} &= m_{k}\frac{y}{m+y}e^{-\beta h},
\end{align}
therefore
\begin{equation}
m_{k}=m_{2}q^{k-2}
\end{equation}
where
\begin{equation}
q=\frac{y}{m+y}e^{-\beta h}.
\end{equation}
The number of all DWs can be expressed as
\begin{equation}
m=\sum_{k=2}^{\infty}m_{k}=m_{2}\frac{1}{1-q},
\end{equation}
therefore we can expressed all $m_{k}$ and $y$ as
\begin{equation}
m_{k}=m(1-q)q^{k-2},\quad
y=1-2n-3m-\frac{1}{1-q}m.
\end{equation}
At this point we only have five parameters $(\alpha,n,m,q,y)$ which
satisfy the following system of equations
\begin{align}
& \int_{0}^{\pi}\frac{dp}{\pi}\frac{1}{e^{\beta(e(p)-\alpha)}+1} =\frac{n}{1-n-2m}\label{eq:eqsys1-1},\\
& \alpha =\frac{T}{2}\log\frac{m(1-q)}{y+m}\label{eq:eqsys2-1},\\
&\nonumber \int_{0}^{\pi}\frac{dp}{\pi}\log(1+e^{-\beta(e(p)-\alpha)}) \\
&\qquad\qquad=-\beta\alpha+2\log\frac{y}{m+y}-\beta h,\label{eq:eqsys3-1}\\
& q =\frac{y}{m+y}e^{-\beta h}, \label{eq:eqsys4-1}\\\
& y =1-2n-3m-\frac{1}{1-q}m. \label{eq:eqsys5-1}\
\end{align}

Let check the $h\to0$ limit. In this limit the equations (\ref{eq:eqsys4-1},\ref{eq:eqsys5-1})
simplify as
\begin{equation}
q=\frac{y}{m+y}\quad\longrightarrow\quad
1-q=\frac{m}{m+y},\quad
y=\frac{1}{2}(1-2n-4m).
\end{equation}
Substituting (\ref{eq:eqsys2-1}) and (\ref{eq:eqsys3-1}) we obtain
the $h=0$ equations (\ref{eq:eqsys2}),(\ref{eq:eqsys3})
exactly.

\begin{figure*}[ht]
\begin{centering}
\includegraphics[width=0.3\textwidth]{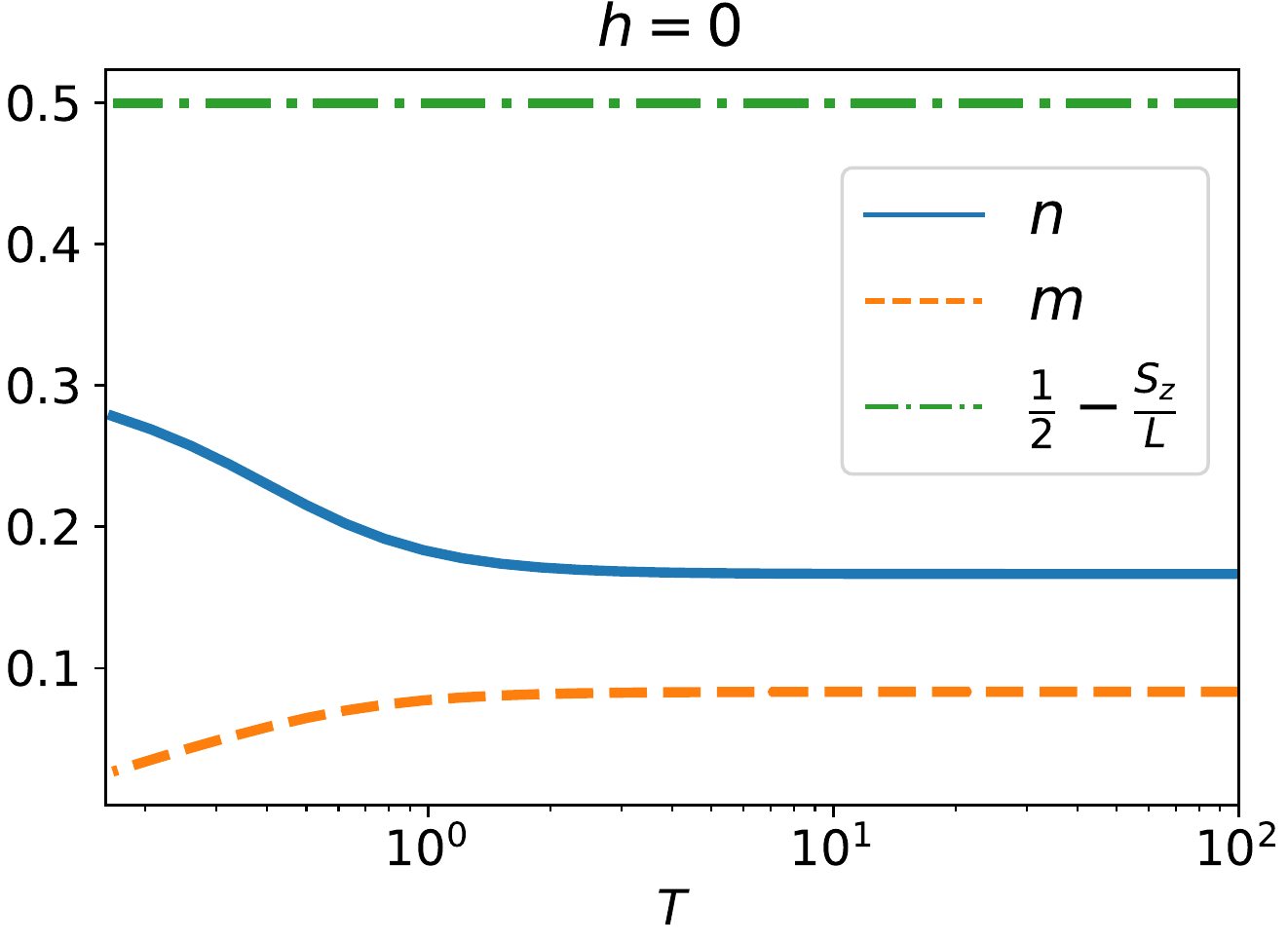}\hspace{0.04\textwidth}\includegraphics[width=0.3\textwidth]{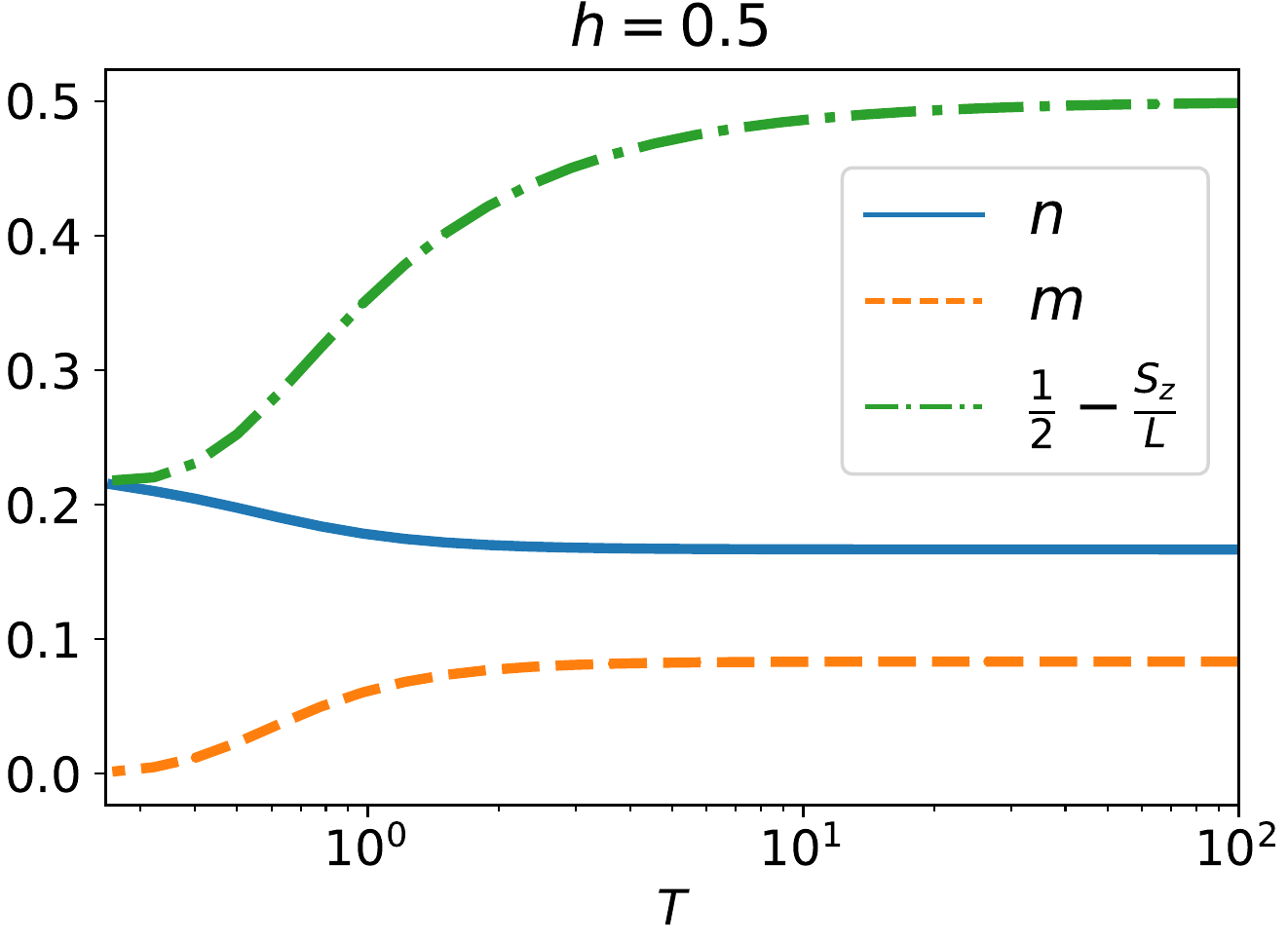}\hspace{0.04\textwidth}\includegraphics[width=0.3\textwidth]{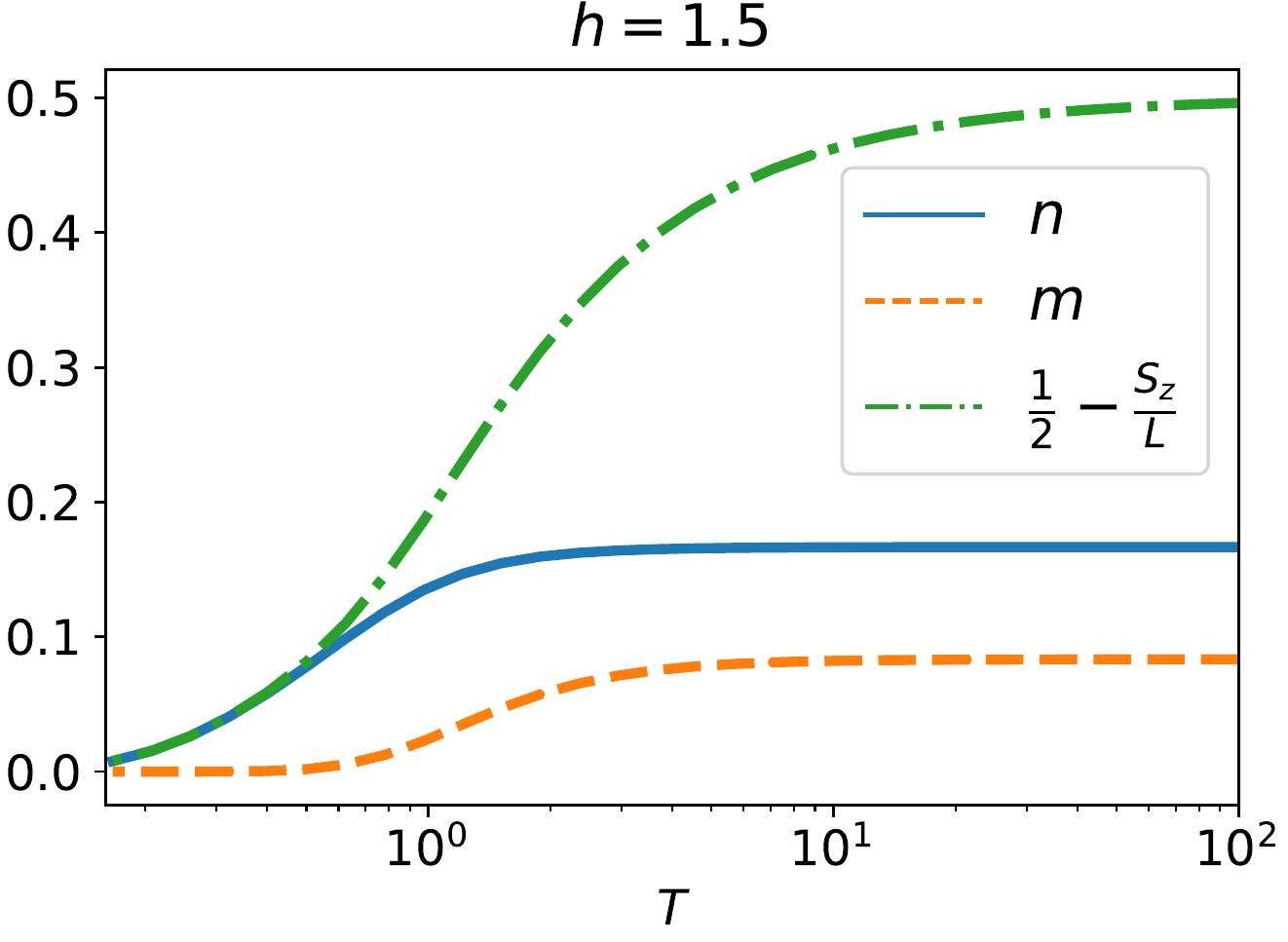}
\par\end{centering}
\caption{Finite temperature states for $h=0,0.5,1.5$. The orange, blue and
green dots show the values of $m$, $n$ and $(L/2-S_{z})/L$ with
respect to the temperature.}
\label{fig:finiteTh}
\end{figure*}

Figure \ref{fig:finiteTh} shows the numerical results for various $h$.
The blue, orange and green curves are the values of $n$, $m$ and $(L/2-S_{z})/L$ w.r.t. the temperature.
The graph on the left shows the $h=0$ case which was already plotted on Figure \ref{fig:TBAopen}, but now we are able to calculate the total spin  for the thermal states.
We can see that the total spin is zero for every temperature.
In the second plot we can see the $h=0.5$ case.
The low temperature limit agrees with ground state analysis i.e. the ground
state is characterized by a finite particle density $n_0\geq 0$.
The third picture shows the $h=1.5$ case.
Now we can see that the $n$ and the total spin goes to $0$ and $L/2$ for zero temperature as we expected from the ground state analysis.
We can also see that the DW density $m$ goes to zero at $T\to0$ and the $T\to\infty$ limit agrees for all $h$.

The equations are simplified again if we use only the variable $x$.
Now we define it as
\begin{equation}
 x^2 = e^{-2\beta \alpha} = \frac{y+m}{m(1-q)}.
\end{equation}
Since
\begin{equation}
 \frac{y}{m+y} = 1 - \frac{x^{-2}}{1-q},
\end{equation}
the equation \eqref{eq:eqsys3-1} can be written as
\begin{equation}
\label{eq:nzeroTBA}
\int_{0}^{\pi}\frac{dp}{\pi}\log(x+e^{-\beta e(p)}) =
2\log F_h(x),
\end{equation}
where
\begin{equation} \label{eq:Fdef}
F_h(x) = \left(x -\frac{x^{-1}}{1-q} \right) e^{-\beta h/2}.
\end{equation}
The variable $q$ can be expressed from \eqref{eq:eqsys4-1} which reads as
\begin{equation}
q = \left(1 -\frac{x^{-2}}{1-q} \right)e^{-\beta h},
\end{equation}
therefore
\begin{equation}
 1-q = \frac{  x\sinh\left(\frac{\beta h}{2}\right) +
               \sqrt{1+x^2\sinh^2\left(\frac{\beta h}{2}\right)} }{x}
      e^{-\beta h/2}.
\end{equation}
Substituting \eqref{eq:Fdef} we can obtain that
\begin{equation}
\label{eq:nzeroFh}
 F_h(x)=\frac{x^2-1}{\sqrt{1+x^2\sinh^2\left(\tfrac{\beta h}{2}\right)}+x\,\cosh\left(\tfrac{\beta h}{2}\right)}.
\end{equation}

The free energy can be also expressed with $x$ as
\begin{equation}
	f/T = \beta h -\log \left(x -\frac{x^{-1}}{1-q} \right) =
	\beta h/2 - \log F_h(x).
\end{equation}

\subsection{Thermodynamics in the periodic case}

An alternative way is to consider the periodic model and to derive the
Thermodynamic Bethe Ansatz (TBA) equations based on the Bethe
equations \eqref{generalBA}. Then we obtain a set of non-linear
integral equations, as usually in the TBA framework \cite{Takahashi-Book}. However, the
present case is different from the situation in the XXZ chain, because
the scattering kernels are constants. This leads to rather simple
equations as opposed to the full XXZ problem.

In the thermodynamic limit we consider again the Bethe root and hole densities
$\rho_n(p)$, $\rho_{\text{h},n}(p)$, but now with the usual normalization such that in volume $L$ the number of
particles/holes for $n$-strings between $p$ and $p+dp$ is $L\rho_n(p)dp$ and $L\rho_{n,h}(p)dp$.
The densities satisfy
\begin{multline}
\label{rhorh2}
\rho_{k}(p)+\rho_{\text{h},k}(p)= \delta_{k,1}\\
+\int_{-\pi}^\pi \frac{dp'}{4\pi}
 \left(\rho_{\text{h},k-1}(p')+\rho_{\text{h},k+1}(p')\right).
\end{multline}
This equation can be derived from  \eqref{generalBA} using the same
steps that lead to similar decoupled equations in the XXZ chain for
$\Delta>1$, see \cite{Takahashi-Book}. We see that the total densities
are constants, because the r.h.s. of these equations does not carry a $p$-dependence.

In the finite temperature situation we introduce the so-called
$Y$-functions as $Y_n=\rho_{\text{h},n}/\rho_n$. They satisfy the TBA equations
\begin{multline}
  \log Y_n(p)=\beta\cos(p)\delta_{n,1}\\
  +\int_{-\pi}^{\pi} \frac{dp'}{4\pi} \log(1+Y_{n-1}(p'))(1+Y_{n+1}(p')).
\end{multline}
It is easy to see that only $Y_1(p)$ is a non-trivial function of $p$
all the rest of the $Y$-functions are constants. Written more
explicitly, we have the set of equations
\begin{align}
\nonumber
&2\log Y_1(p)=2\beta\cos(p)+\log(1+Y_2),\\
&2\log Y_2=\int_{-\pi}^{\pi}\frac{dp'}{2\pi}  \log(1+Y_1(p'))+\log(1+Y_3),\\\nonumber
&Y_n^2=(1+Y_{n-1})(1+Y_{n+1}),\qquad n\ge 3.
\end{align}
The TBA equations can be solved when combining with the asymptotics of $Y$-functions in the large $n$ limit.
For $h>0$ the asymptotics is given by
\begin{align}
\label{eq:asymY}
\lim_{n\to \infty}\frac{\log Y_n}{n}=\beta h.
\end{align}
This implies that for $n\to\infty$, $Y_n$ grows exponentially in $n$ as $Y_n\sim e^{\beta h n}$.
At $h=0$ the asymptotics of the $Y_n$ for large $n$ grows polynomially in $n$.
Finally, the free energy %(\ref{eq:freeEnergy})
is given by
\begin{align}
\frac{F}{L}=\frac{h}{2}-T\int_{-\pi}^{\pi}\frac{dp}{4\pi} \log(1+Y_1(p)).
\end{align}
We now discuss the solution of TBA.

Let us denote
\begin{align}
\label{eq:Y2mu}
\beta\alpha=-\frac{1}{2}\log(1+Y_2).
\end{align}
The TBA equation can be written as
\begin{align}
\label{eq:TBAtoSolv}
&2\log Y_2=\int_{-\pi}^{\pi}\frac{dp'}{2\pi}\log\left(1+e^{\beta (\cos(p')-\alpha)}\right)+\log(1+Y_3),\\\nonumber
&Y_n^2=(1+Y_{n-1})(1+Y_{n+1}),\qquad n\ge 3.
\end{align}
We first consider the second equation in (\ref{eq:TBAtoSolv}), which is a second order difference equation. The general solution to this equation is given by
\begin{align}
Y_n=\left(\frac{\sinh([y_0+n]a_0)}{\sinh(a_0)}\right)^2-1,
\end{align}
where $a_0$ and $y_0$ are two constants. Using the asymptotics (\ref{eq:asymY}), we can fix $a_0=\beta h/2$. To fix $y_0$, we consider (\ref{eq:Y2mu}), which can be written as
\begin{align}
Y_2=e^{-2\beta\alpha}-1=\left(\frac{\sinh([y_0+n]\beta h/2)}{\sinh(\beta h/2)}\right)^2-1.
\end{align}
From this equation, we can solve $y_0$ in terms of $\alpha$. Plugging back to the first equation of (\ref{eq:TBAtoSolv}), we obtain the following equation for the variable $x=e^{-\beta\alpha}$
\begin{align}
\label{eq:xeq}
2\log F_h(x)=
\int_{-\pi}^{\pi}\frac{dp'}{2\pi}  \log\left(1+x\,e^{\beta\cos(p')}\right),
\end{align}
where
\begin{align}
F_h(x)=\frac{x^2-1}{\sqrt{1+x^2\sinh^2\left(\tfrac{\beta h}{2}\right)}+x\,\cosh\left(\tfrac{\beta h}{2}\right)}.
\end{align}
This result is the same as in (\ref{eq:nzeroFh}). The TBA equation (\ref{eq:xeq}) coincides precisely with the one
(\ref{eq:nzeroTBA}) derived in the previous section. Although (\ref{eq:nzeroTBA}) is derived by considering open
boundary condition, the differences due to boundary conditions are negligible in the thermodynamic limit. Thus we
obtained the same TBA equation as expected.

%For fixed $\beta$ and $h$, the equation (\ref{eq:xeq}) can be solved straightforwardly by iteration.\par

%The $h=0$ case can be obtained simply by taking $h=0$ in (\ref{eq:xeq}). We have
%\begin{align}
%F_0(x)=x-1.
%\end{align}
%The constant $Y$-functions are given by
%\begin{align}
%Y_n=(n+y_0)^2-1,\qquad y_0=x-2.
%\end{align}
%It is easy to see that for $h=0$, the TBA equation
%\begin{align}
%\label{eq:xeqh0}
%2\log F_0(x)=
%\int_{-\pi}^{\pi}\frac{dp'}{2\pi}  \log\left(1+x\,e^{\beta\cos(p')}\right)
%\end{align}
%is precisely the one derived in the previous subsection (...), which has been solved numerically.

\subsection{Thermodynamics in the bond picture}

We can also compute the thermodynamics using the Bethe equations \eqref{bondBetheeq}. A direct application of the TBA
framework would lead to two coupled integral equations, corresponding to the two particle species. However, instead of
copying the standard formulas of TBA we can also derive the result using a simple reasoning and formulas known from free
theories.

Our first argument is that in the direct equations \eqref{bondBetheeq} we can assume that the total momentum of the
particles and the DW's is zero; the thermodynamical computations are not sensitive to this assumption. Then we are faced
with quantization conditions for two free models, such that the energy is computed only from the particles but it is
insensitive to the DW's. Our strategy is similar as in the previous subsection: we compute the free energy for arbitrary
overall densities $n$ and $\tilde m$ of the particles and DW's and afterwards we perform a saddle point analysis. For a
direct comparison with the previous formulas we will use $\tilde m=2m$, because in the earlier computations $M$ and $m$
denoted the total number and density of the bound states, whereas here $\tilde m$ stands for the total density of the
DW's. Each bound state has two domain walls at the two ends, thus the factor of 2 in the relation.

The free energy density will be a sum of the particle and DW contributions:
\begin{equation}
  f=f_p+f_{DW}.
\end{equation}
The computation of the free energy associated to the DW's is rather simple: as there is no energy, the free energy
contribution comes only from the entropy, which is computed easily. We find
\begin{multline}
  \beta f_{DW}= -(1-2n-\tilde m) \log (1-2n-\tilde m)\\
  +(1-2n-2 \tilde m)\log (1-2n-2 \tilde m) +\tilde m \log  \tilde m.
\end{multline}
The free energy of the particles at fixed particle number $n$ is computed using a Lagrange multiplier $\alpha$:
\begin{equation}
   \beta   f_p=- (1-n-\tilde m)  \int_{-\pi}^\pi \frac{dp}{2\pi} \log(1+e^{-\beta (e(p)-\alpha)})+\beta\alpha n.\\
\end{equation}
The value of $\alpha$ is then found from
\begin{equation}
  \frac{\partial f}{\partial \alpha}=0
\end{equation}
leading to the condition
\begin{equation}
   (1-n-\tilde m)  \int_{-\pi}^\pi \frac{dp}{2\pi} \frac{1}{1+e^{\beta (e(p)-\alpha)}}=n.
\end{equation}
This equation is the same as \eqref{eq:eqsys1} after the appropriate re-normalization $\tilde m=2 m$.

\section{Solvable Quench Dynamics}

The model given by \eqref{H} is special because in certain cases the real time dynamics can be solved exactly, leading to exact formulas for the time
dependence of local observables. The techniques to be used are
completely analogous to those used in \cite{sajat-qboson,sajat-q2}, and
they were originally developed in \cite{bog-ising-limit,bog-ising-limit2,qboson-bog1,qboson-bog2,qboson-bog3}.

Let us focus on quench problems, where we prepare the system in an initial state $\ket{\Psi_0}$ and let it evolve with
the model Hamiltonian. The simplest problems are those which involve only the particle excitations and not the Domain
Walls. Such a situation arises if $\ket{\Psi_0}$ has strictly zero overlap with states including DW's. It is relatively
easy to construct such initial states: the only requirement is that in the computational basis there can not be two or
more spin excitations on neighbouring positions. For such initial states the overlaps can be computed as a sum over
determinants  using  the exact wave function \eqref{Bethestate}. The simplest cases are those when the initial state is
an element of the computational basis, such that the overlaps are single determinants given by \eqref{Bethestate}. We
put forward that interesting physics arises in those cases where the bound states are also present; this is discussed
 in the next Section.

The simplest state with zero overlap with bound states is probably the N\'eel state where there is a particle excitation at
every second site. Using the notation
\eqref{xnot} the state is given by
\begin{equation}
  \ket{\Psi_0}=\ket{2,4,6,\dots}.
\end{equation}
However, in this model the N\'eel state is actually an eigenstate with eigenvalue 0.

The next simplest case is the period 3 state
\begin{equation}
  \ket{\Psi'_0}=\ket{3,6,9,\dots}.
\end{equation}
Here we have $N$ particles in a volume $L=3N$. This state is not an eigenstate and the overlaps with the Bethe states
are given by
\begin{equation}
\label{pp-psi}
  \skalarszorzat{\Psi'_0}{\pp}=\det_{jk} e^{ip_j (2k+1)}.
\end{equation}
This follows from \eqref{Bethestate} after the substitution $x_k=3k$. We will see that the treatment of the quench from
this state is very much analogous to the quench problem treated in Section 6 of \cite{sajat-qboson}.

For simplicity we restrict ourselves to the zero momentum sector,
therefore we consider the initial state
\begin{equation}
  \label{psinulldef}
  \ket{\Psi_0}=\frac{1+U+U^2}{\sqrt{3}}\ket{\Psi_0'},
\end{equation}
where $U$ is the one-site cyclic shift operator.

In this case the overlaps can be expressed simply as
\begin{equation}
\label{psin}
|\langle\Psi_0|\pp\rangle|^2=\prod_{j<k}|e^{i2p_j}-e^{i2p_k}|^{-2}.
\end{equation}
This overlap will be evaluated in the following.

In the zero momentum sector the Bethe equations are
\begin{equation}
e^{i2Np_j}=-1,
\end{equation}
where we used $L=3N$ and we assumed an even $N$ for simplicity.

Solutions are given by
\begin{equation*}
  p_j=\frac{\pi (2I_j-1)}{2N},\qquad I_j=1,2,\dots 2N.
\end{equation*}
The zero momentum Bethe states are given by the subsets
\begin{equation*}
  \{e^{ip_j}\}_L \subset \{\omega\}_{2N}, \quad \omega_k=e^{i\frac{\pi (2k-1)}{2N}},\quad k=1,2,\dots 2N
\end{equation*}
satisfying the constraint $\prod_j e^{ip_j}=1$.
The numbers $\omega_k$ can be paired such that
\begin{equation*}
  \{\omega\}_{2L}=\{(\omega_k,-\omega_k)\}_{k=1\dots N}.
\end{equation*}
It follows from formula \eqref{psin} that
the overlap is non-vanishing only if exactly one rapidity is chosen from
each pair. Therefore, the states with non-vanishing overlap are given by
\begin{equation}
\label{ezek}
  a_j=s_j \omega_j,\quad\text{where}\quad s_j=\pm 1,\quad j=1\dots N,
\end{equation}
with the constraint that the total momentum is zero. We have
\begin{equation}
\label{zerom}
1=  \prod_{j=1}^N e^{ip_j}=\prod_{j=1}^N (s_j \omega_j)=e^{i\pi N/2}\prod_{j=1}^N s_j.
\end{equation}
We assumed that $N$ is even and the zero momentum constraint implies that there are a total number of $2^{N-1}$
states with non-vanishing overlap.

It can then be seen that the overlaps \eqref{psin} are all equal and it was shown in \cite{sajat-qboson} that their value is
\begin{equation}
  |\langle\Psi_0|\pp\rangle|^2=N^N.
\end{equation}

Our goal is to compute the time evolution of local observables $\ordo$ using the spectral expansion
\begin{equation}
\label{sum}
\vev{\ordo(t)}=\sum_{\pp,\kk}
\frac{\skalarszorzat{\Psi_0}{\pp}\bra{\pp}\ordo\ket{\kk}\skalarszorzat{\kk}{\Psi_0}}
{  \skalarszorzat{\pp}{\pp}  \skalarszorzat{\kk}{\kk}}
  e^{-i(E_{\bf k}-E_{\bf p})t}.
\end{equation}

We focus on simple local observables, with our main candidates being the operators measuring the emptiness formation
probability (EFP). We define the $\ell$-site local operator $\emp_\ell(x)$ positioned at $x$ as
\begin{equation}
  \emp_\ell(x)=\prod_{j=1}^\ell \frac{1+\sigma^z_{x-1+j}}{2}.
\end{equation}
We also define their space average:
\begin{equation}
 \bar \emp_\ell=\frac{1}{L}\sum_{j=1}^L  \emp_\ell(j).
\end{equation}

The operators $\bar \emp_1$ and $\bar\emp_2$ are conserved, because
they are linear combinations of the charges $Q_1$, $Q_2$ and the
identity. Thus the first member of the series with non-trivial time
evolution is $\bar\emp_3$.

The matrix element of a single operator $\langle
\pp|\emp_{\ell}(x)|\kk\rangle$ for arbitrary $\ell$ was computed in \cite{bog-ising-limit,bog-ising-limit2}
\begin{equation}
\label{EFP}
\langle \pp|\emp_{\ell}(x)|\kk\rangle=\prod_{j\le k}\frac{1}{\left(e^{i\uc_j}-e^{i\uc_k}\right)\left(e^{i\ub_j}-e^{i\ub_k}\right)}
\det\mathcal T,
\end{equation}
where
\begin{equation}
\mathcal T_{jk}= \frac{1-e^{i(\ub_j-\uc_k)(L-\ell)}e^{i\left((N-1)(\uc_k-\ub_j)\right)}}{e^{i\ub_j}-e^{i\uc_k}}.
\end{equation}
%
%and  $\{\uc\}$ and $\{\ub\}$ are rapidities that parametrize $\kk$ and $\pp$ states correspondingly.

%
The norm of the Bethe states is given by (see \cite{pronko-abarenkova-ising-limit})
\begin{equation}
\label{norm}
  \skalarszorzat{\pp}{\pp}=L(L-N)^{N-1}.
\end{equation}

The matrix elements of $\mathcal{T}$ can then be rewritten as
\begin{align}
&\mathcal T_{jj}=(N-L+\ell-1)e^{-2ik_j},\quad \uc_j=\ub_k,\\
&\mathcal T_{jk}=e^{i(\ell-1)(\ub_j+\uc_k)}\frac{\sin\left((\ell-1)(\ub_j-\uc_k)\right)}{\sin(\ub_j-\uc_k)},\quad \uc_j\ne \ub_k.
\end{align}
It is easy to see that for diagonal part of $\mathcal T$ the rank is
given by
the number of coinciding elements in the sets $\buc$, $\bub$.

For $\ell=1$ the non-diagonal part is equal to zero, thus the only
non-zero contribution to the sum \eqref{sum} is the case $\buc=\bub$
and we obtain trivial time dependence.
 In the case $\ell=2$ the non-diagonal part has rank 1,
thus it is possible to have a non-zero contribution to \eqref{sum} in a case the diagonal part of matrix $\mathcal T$ has the rank at least $N-1$.
This happens if there are $N-1$ rapidities $\uc_j$  coinciding with some of rapidities from set $\bub$.
However, due to the $P=0$ condition it is not possible to have a case where the two sets differ only by one rapidity.
Thus, for $\ell=2$ case the selection rule for the form factors is
again $\buc=\bub$ and we  have only the static
contribution. These findings are consistent with the fact that
$\emp_1$ and $\emp_2$ are related to the conserved charges $Q_1$ and $Q_2$.

Finally, for $\ell=3$  the non-diagonal part of $\mathcal T$ has rank $2$, thus in order to have a non-zero determinant
the rank of the diagonal part should be at least $N-2$.
Then there is a time dependence of $\emp_{3}(x)$ expectation value if two rapidities are different in the sets $\buc$ and $\bub$.

The further computation of dynamics for the case $\ell=3$ is absolutely similar to the one performed in \cite{sajat-qboson}
for the q-boson model. Thus we omit simple
details and give here the final result
\begin{multline}
  \label{qfi1}
\langle\psi(t)|\emp_{3}(x)|\psi(t)\rangle=\frac{1}{6}-\frac{1}{6}\left(\frac{1}{N}\sum_{a}\cos(2\cos(c_a)t)\right)^2\\
-\frac{1}{6}\left|\frac{1}{N}\sum_{a}\sin(2\cos(c_a)t)e^{ic_a}\right|^2,
\end{multline}
where $c_a=\pi (2a-1)/(2N)$, $a=1,\dots,N$.
In the thermodynamic limit the last expression  can be presented as
\begin{multline}
\langle\psi(t)|\emp_{3}(x)|\psi(t)\rangle
=\frac{1}{6}-\frac{1}{6}\left(\int_0^\pi\frac{dz}{\pi}\cos(2\cos(z)t)\right)^2\\
-\frac{1}{6}\left|\int_0^\pi \frac{dz}{\pi}\sin(2\cos(z)t)e^{iz}\right|^2.
\end{multline}
The only difference with the corresponding formula of  \cite{sajat-qboson}
(see eq. (6.16) of that work) is the appearance of the extra factor of $1/3$, which can be traced back to the difference
in the norm \eqref{norm} and an overall factor of 3 originating in the definition \eqref{psinulldef} of the initial state.

The integrals above actually describe the Bessel functions of the first kind, so we can write
\begin{equation}
  \langle\psi(t)|\emp_{3}(x)|\psi(t)\rangle
=\frac{1}{6}\left[1-(J_0(2t))^2-(J_1(2t))^2\right].
\end{equation}

The asymptotic behaviour for large $t$ is
\begin{equation}
  \langle\psi(t)|\emp_{3}(x)|\psi(t)\rangle=\frac{1}{6}\left[1-\frac{1}{\pi t}+\ordo(t^{-2})\right].
\end{equation}

We compared \eqref{qfi1} to results from exact diagonalization and found complete agreement.

\section{Breakdown of the GGE: Persistent oscillations}

The large number of degeneracies in the spectrum has an interesting consequence: it can lead to the appearance of persistent
oscillations in certain quench problems. Such persistent oscillations can appear in $SU(2)$-symmetric spin chains if a magnetic
field is applied, or in more complicated situations \cite{time-crystal} where the degeneracies have a somewhat
unexpected algebraic origin 
\cite{korff-degeneracies,semisyclic,time-crystal}. We will argue below that the present model also supports persistent oscillations. However, in
contrast to the solvable quench treated in the previous Section, the treatment of the oscillations is much more involved
from a theoretical point of view. At present we do not have an exact derivation of the oscillations, therefore we
present their existence as a conjecture, which we support by numerical evidence.

In order to understand the effect we briefly discuss the process of
equilibration in integrable models. Let us again assume that we release the model from an initial state $\ket{\Psi_0}$
and we investigate the real time
dynamics of local observables $\ordo(t)$. Inserting complete sets of states $\ket{a}$ and $\ket{b}$ in finite volume we
get
\begin{equation}
  \label{doublesum}
  \vev{\ordo(t)}=\sum_{a,b} \skalarszorzat{\Psi_0}{a}\bra{a}\ordo\ket{b}\skalarszorzat{b}{\Psi_0} e^{-i(E_b-E_a)t}.
\end{equation}
In the long time limit (or for long time averages) it is usually assumed that this double sum can be constrained to the
diagonal contributions. The idea is that in the generic situation there are not too many degeneracies in the
spectrum, the energy levels $E_a$ can be considered sufficiently independent, and for large enough times the
phases of the off-diagonal contributions average out to zero. This leads to the Diagonal Ensemble (DE)
\begin{equation}
\lim_{T\to\infty}\frac{1}{T} \int_{t=0}^T dt\ \vev{\ordo(t)}=\sum_a \left|\skalarszorzat{\Psi_0}{a}\right|^2 \bra{a}\ordo\ket{a}.
\end{equation}
From this expression it is then argued, that the mean values are given
 by the Generalized Gibbs Ensemble (GGE), which incorporates all conserved
charges of the model \cite{essler-fagotti-quench-review}. The derivation of the GGE is possible if the so-called
Generalized Eigenstate Thermalization Hypothesis (GETH) is satisfied. If these assumptions hold then the system will  equilibrate and the steady states
will not support oscillations in any local observable.

The situation is different in the presence of a large number of degeneracies.
Let us now denote by $E_a$ the energy eigenvalues, and by a further discrete index $j$ the states $\ket{a,j}$ in the
degenerate eigenspaces. Then we have the double sum
\begin{equation}
  \label{doublesum2}
\lim_{t\to\infty} \vev{\ordo(t)}=\sum_{a}\sum_{j,k} \skalarszorzat{\Psi_0}{a,j}\bra{a,j}\ordo\ket{a,k}\skalarszorzat{a,k}{\Psi_0}.
\end{equation}
In such a situation the long time limit is given not only by the mean values, but rather by a complicated sum involving
many off-diagonal matrix elements. Generally we can expect the breakdown of the GGE in such a case. Furthermore, if
the degeneracies can be lifted such that the energy differences remain commensurable, then we can observe persistent
oscillations.

Let us consider concrete examples in our model. We perform quenches with the Hamiltonian \eqref{H}, where we set
$\mu=0$ for simplicity, but we consider the $h=0$ and $h\ne 0$ cases as well. The larges number of degeneracies appear
at $h=0$, while some of these are broken up with fixed differences if $h$ is switched on.

For the initial state our main candidate will be  the ferromagnetic state in the $x$ direction, which is given by
\begin{equation}
  \ket{\Psi_0}=\otimes_{j=1}^L \frac{1}{\sqrt{2}}
  \begin{pmatrix}
    1 \\ 1
  \end{pmatrix}.\label{initial_state}
\end{equation}
This is a good choice because it excites various combinations of particles and domain walls. A key property is that the
state breaks the $U(1)$-symmetry of the Hamiltonian, and after the quench the system will be populated by states with
different values of the $S^z$ global charge.

We choose local operators which also break this $U(1)$-symmetry.
We look specifically at
\begin{equation}
D_k=\prod_{j=1}^k \sigma^-_{j}.
\end{equation}
These operators change the global $S^z$ charge. Therefore their mean values in the DE (or in the GGE) are identically
zero.

We conjecture that in our model these observables either tend to a finite mean value (for $h=0$) or they display
oscillations (for $h\ne 0$). Both behaviour signals
the breakdown of the GGE and the presence of the off-diagonal contributions in \eqref{doublesum2}.
The difference between the two behaviours is explained easily: if $h$ is switched on then some of degeneracies are split
with fixed amounts according to the global charge $S^z$. The operator $\ordo=D_k$ changes $S^z$ by a fixed
amount, thus every non-zero contribution in the double sum \eqref{doublesum2} will receive the same phase $e^{ihkt}$.

\subsection{Numerical results}

\begin{figure*}[ht]
\centering
\subfigure[]{
\includegraphics[scale=0.62]{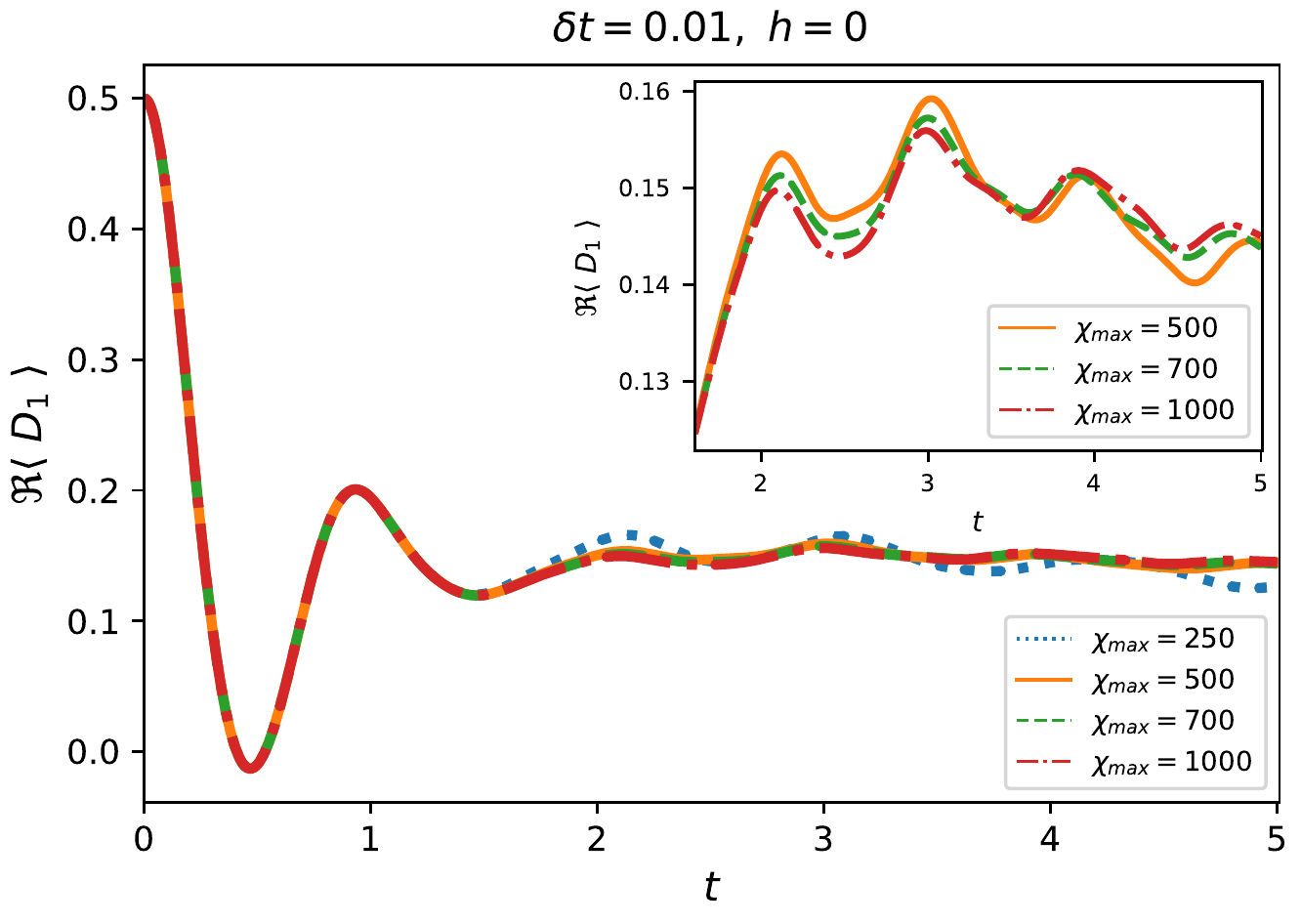}}
\quad
\subfigure[]{%
\includegraphics[scale=0.62]{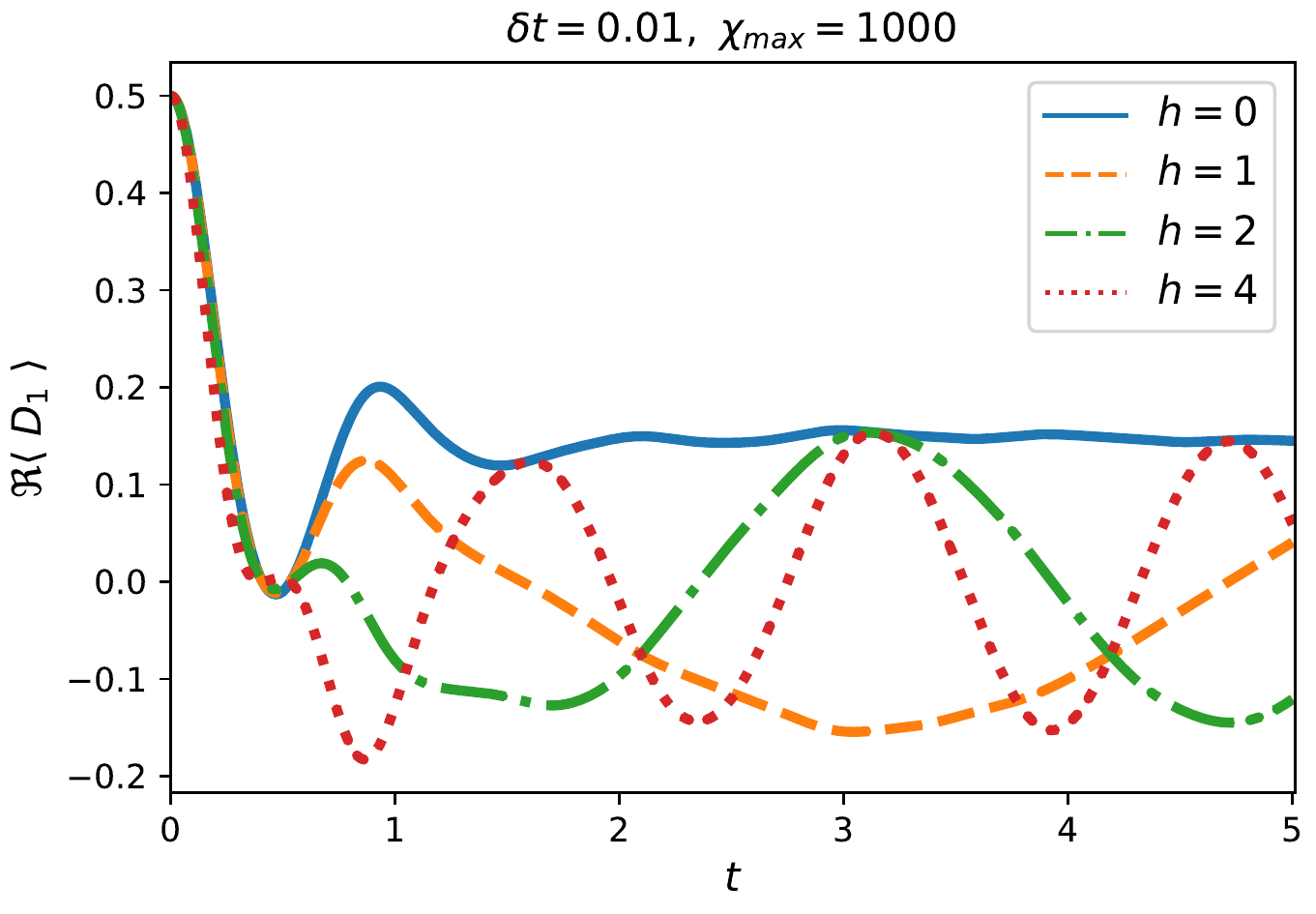}}

\caption{(a) The real part of the expectation value of $D_1$ for $h=0$ with different maximal bond dimensions. On the
  inset the same curves are magnified for $t>1.6$. The data seems to confirm a non-vanishing asymptotic value, but we
  believe that this evidence is not decisive. The observation of longer times would be desirable, which is not possible
  with our numerical implementation. (b) The real part of the expectation value of $D_1$ for different values of
  $h$. The maximal bond dimension is $\chi_{max}=1000$, the Trotter time step is $\delta t=0.01$. We observe the predicted
  oscillation of the expectation value with a frequency directly given by $h$.}
\label{fig:iTEBD1}
\end{figure*}

We support our conjecture with numerical evidence, obtained from iTEBD \cite{vidal-itebd0,vidal-itebd1} simulations. We
used the example code in 
\cite{pollmann-notes} as a starting point and modified it to our purposes to simulate real-time evolution governed by a
four-site Hamiltonian: the state of the system is represented as a four-site translational invariant matrix product
state (MPS) \cite{mps-intro1} with four sets of matrices
\begin{equation}
\begin{split}
\ket{\Psi}=\!\!\!\!\!\!\sum_{\dots,j_k,j_{k+1},j_{k+2},j_{k+3},\dots}&\!\!\!\!\!\!\!\!\!\!\!\!\!\!\!\dots\Lambda_0\Gamma_0^{j_k}\Lambda_1\Gamma_1^{j_{k+1}}\Lambda_2\Gamma_2^{j_{k+2}}\Lambda_3\Gamma_3^{j_{k+3}}\dots \\ &\times\ket{\dots , j_k,j_{k+1},j_{k+2},j_{k+3},\dots}, \label{MPS_state}
\end{split}
\end{equation}
where both the $\Gamma_k^j$ and $\Lambda_k (k=0,1,2,3)$ are $\chi\times\chi$ matrices, where $\chi$ is the so-called
bond dimension. The latter are diagonal and they
contain the singular values corresponding to the bi-partition of the system at the given bond. The algorithm uses a first
order Suzuki-Trotter decomposition for the time evolution operator. We initialize our system in the state $\ket{\Psi_0}$
given by \eqref{initial_state} (since it is a simple product state, it can be easily written in a form like
\eqref{MPS_state}) and evolve it with a Trotter step of $\delta t=0.01$. Unfortunately the entanglement entropy grows
rapidly in time, which limits the applicability of the method to short times. To check the validity of our results we
use several different maximal bond dimensions $\chi_{max}$. The results are shown in Figure~\ref{fig:iTEBD1}. The curves
confirm the theoretical expectations in the investigated time frame.

\section{Relation with $T\bar T$-deformations}

\label{sec:ttbar}

Here we point out an interesting relation with the so-called $T\bar T$-deformations of QFT's. Such deformations can be
introduced for any QFT, and for the integrable cases it is known that they preserve the integrability of the model
\cite{Zam-TTbar,Smirnov-Zam-TTbar,roberto-istvan-ttbar}. The $T\bar T$-deformation modifies the scattering matrix of the
model, which is well understood and which we discuss below. Analogues of the $T\bar T$-deformation for spin chains were
discussed in \cite{sajat-ttbar,sfondrini-ttbar}
pointing out that these transformations are essentially the same as a certain type of long-range deformation studied
earlier in the context of the AdS/CFT correspondence \cite{beisert-long-range-2}.

A $T\bar T$-like deformation can be introduced for any pair of extensive conserved quantities $Q_\alpha$ and
$Q_\beta$. To first order the deformation consists in modifying the Hamiltonian $H$ of a given model as
\begin{multline}
H'=H\\
+\kappa \left(  J_\alpha(x)q_\beta(x)-q_\alpha(y) J_\beta(x+1)\right)+\ordo(\kappa^2),
\end{multline}
where $q_{\alpha}(x)$ are the charge densities and $J_{\alpha}(x)$ are the current operators describing the
flow of the charges \cite{sajat-ttbar,sfondrini-ttbar}. Here we gave the formula for the spin chain situation, but
an analogous expression gives the corresponding perturbation in the QFT case as well
\cite{Zam-TTbar,Smirnov-Zam-TTbar,roberto-istvan-ttbar}.

If $S(p,k)=e^{i\delta(p,k)}$ is the two-particle scattering matrix of the model, then under the perturbation above it gets deformed as
\begin{multline}
 \delta'(p,k)= \delta(p,k)\\
 + \kappa(h_\alpha(p)h_\beta(k)-h_\alpha(k)h_\beta(p))+\ordo(\kappa^2)
\end{multline}
where $h_{\alpha,\beta}$ are the one-particle eigenvalue functions of the charges $Q_\alpha$ and $Q_\beta$.

In QFT  the actual $T\bar T$-deformation corresponds to choosing the energy and the momentum as two
charges. An other important case is the so-called hard rod deformation, which corresponds to choosing the particle number and the momentum as the two charges \cite{doyon-cardy-ttbar,yunfeng-hardrod-1}. In this
case the modification of the $S$-matrix is simply
\begin{equation}
  \label{pkdef}
  S(p,k)\to S(p,k) \exp\left\{i\kappa(p-k)\right\}.
\end{equation}
It is important that the parameter $\kappa$ can be varied continuously, and usually it is assumed to be small compared
to some characteristic length scale.

In contrast with the QFT situation it is not possible to construct the $T\bar T$ and hard rod deformations on the lattice.
On a technical level this
happens because on the spin chain there is no momentum operator which would be an extensive local charge. On a physical
level this absence comes simply from the discrete structure of the chain.

Quite interestingly, the $S$-matrices that we encountered in this paper have the structure of \eqref{pkdef} with an
integer $\kappa$ with the undeformed $S$-matrix being $\pm 1$.
For example the $S$-matrix factor \eqref{S11} of the particles in our model
 is of this form, thus it can be considered a hard rod deformation of the free $S$-matrix of the XX
 model with $\kappa=-1$. There are other similar examples for this in the literature. For example the $S$-matrix of
 the phase model (see  \cite{sajat-qboson,sajat-q2}) is given by
 \begin{equation}
   S(p,k)=-e^{i(p-k)}.
 \end{equation}
This is the hard rod deformation of the XX model $S$-matrix $S(p,k)=-1$ with deformation parameter $\kappa=+1$.
Similarly, the $S$-matrix of the constrained XXZ model found in \cite{constrained1,constrained2,constrained3} can
be considered a hard rod deformation of the scattering phases of the XXZ model.

In all of the cases mentioned above there is a non-local transformation which actually connects the deformed model with
its undeformed parent model. Then the additional phase in \eqref{pkdef} can be interpreted as the additional effect of certain
displacements dictated by the non-local transformations. These displacements can be interpreted as the particles having
a finite width. Thus we can interpret these models and their non-local transformations as concrete and explicit examples
of the hard rod deformation discussed \cite{doyon-cardy-ttbar,yunfeng-hardrod-1}.

It is known that in QFT the actual $T\bar T$-deformation can be also be interpreted by non-local and field-dependent
transformations, and that the particles can be seen as having acquired a (momentum dependent) finite width
\cite{ttbar-geom1,ttbar-geom2,doyon-cardy-ttbar,yunfeng-hardrod-1}.  The examples discussed so far demonstrate such a
relation in lattice systems.

Regarding the full spectrum of the present model the situation is somewhat more complicated than in the constrained XXZ
model or the phase model. Now the ultimate non-local
transformation that maps the interacting model to an (almost) free model concerns the Maassarani-Mathieu chain, which
is a spin chain with a different local Hilbert space. Thus we have found a new type of non-local map, which nevertheless
belongs to the class of hard rod deformations.

\section{Conclusions}

In this work we studied the so-called folded XXZ model which appeared in the earlier works
\cite{fracton1,folded1,folded2}. The main new results of our work
 are a) the direct connection with the
charges of the XXZ model; b) the various forms of the Bethe Ansatz, both with periodic and open boundary conditions; c)
the three  different descriptions of the thermodynamics; d) the connection with the Maassarani-Mathieu chain;
e) the exact solution for a specific quantum quench problem; f) a conjecture for the presence of persistent oscillations,
supported by numerical proof; g) making a connection to the $T\bar T$ and hard rod deformations, and also to similar
models in the literature.

There are a number of interesting open questions.

First of all, it would be interesting to understand the direct algebraic origin of the present model, and how it fits
into the Quantum Inverse Scattering Approach \cite{Korepin-Book}. Our derivation is capable of producing the charges and
the eigenstates of the model, nevertheless a direct treatment with the Algebraic Bethe Ansatz is not possible. The
reason for this is that the Lax operators which are used in the construction of the transfer matrix of the XXZ model do
not survive the $\Delta\to\infty$ limit, or at least this does not happen in a straightforward way. In fact we can prove
that the model can not be constructed using the standard way, where the Lax operator is chosen as the $R$-matrix in the
fundamental representation: in such cases the charges can be constructed using the boost operator, whereas in our model
$Q_2$ is not dynamical and a direct application of the boost will not produce the next charge $Q_3$; this was discussed
in Section \ref{sec:XXZ}. Thus the model requires a different Lax operator construction, and a direct $\Delta\to\infty$
of the Lax operators of the XXZ chain will not work.

It would be interesting to consider other models with constant scattering lengths, and  hard rod-like $S$-matrix
factors, and to formulate the
common algebraic origin of these models. So far the known models in this family are the phase model (the $q\to\infty$
limit of the $q$-boson model), the Rule 54 model, and the present one, the folded XXZ model. It would be desirable to
develop a unified
algebraic framework for such spin chain models. As a particular case of this problem, it would be
important to uncover relations between the Rule 54 model and the quantum chains that we treated.

The exact results in these relatively simple models could be used to confirm the predictions of Generalized
Hydrodynamics (GHD) in various situations. Such a highly non-trivial check was performed in the recent work
\cite{katja-bruno-rule54-ghd} in the case of the Rule 54 model. We believe that the present spin chain and the related
models are perfect candidates for such checks: they are quantum models that have genuine interactions, nevertheless they
are simple enough that even the spectral sums can be evaluated in certain situations. This presents unique opportunities
for exact computations.

\vspace{0.1cm}
{\bf Acknowledgments}

We are grateful to Maurizio Fagotti, Frank G\"ohmann, Lorenzo Piroli, Toma\v{z} Prosen, Tibor Rakovszky, Roberto Tateo,
Mikl\'os Werner, Lenart Zadnik for useful discussions.
In particular we are thankful to Mikl\'os Werner for his help regarding the iTEBD program code and discussions
concerning the persistent oscillations.

\bibliography{hardrod}

\end{document}